\documentclass{aa}  

\usepackage{graphicx}
\usepackage{txfonts}
\usepackage{amsmath}	% Advanced maths commands
\usepackage{amssymb}	% Extra maths symbols
\usepackage[usenames]{color}
\usepackage{subfig}

\definecolor{Blue}{rgb}{0,0.5,1}
\definecolor{Red}{rgb}{1,0,0}
\definecolor{Orange}{rgb}{1,.38,0}

\begin{document} 
%%%%%%%%%%%%%%%%%%% TITLE PAGE %%%%%%%%%%%%%%%%%%%

% Title of the paper, and the short title which is used in the headers.
% Keep the title short and informative.
    \title{Dense gas formation and destruction in a simulated Perseus-like galaxy cluster with spin-driven black hole feedback}
    \titlerunning{Dense gas in galaxy clusters under the influence of a spin-driven BH jet}

   \author{R. S. Beckmann\inst{1}\thanks{ricarda.beckmann@iap.fr}
          \and
          Y. Dubois\inst{1}
          \and 
          P. Guillard\inst{1}
          \and
          P. Salome\inst{2}
        \and
        V. Olivares\inst{2}
        \and
        F. Polles\inst{2}
        \and 
        C. Cadiou\inst{1}
        \and
        F. Combes\inst{2,3}
        \and 
        S. Hamer \inst{4}
        \and
        M. D. Lehnert\inst{1}
        \and
        G. Pineau des Forets\inst{2}
          }

   \institute{Institut d'Astrophysique de Paris, CNRS UMR 7095,
              Sorbonne Universit\'e, 75014 Paris, France
                \and
                LERMA, Observatoire de Paris, PLS research
                Universit\'e, CNRS, Sorbonne Universit\'e, 75104 Paris, France
                \and
                Coll\`ege de France, 11 Place Marcelin Berthelot, 75005 Paris, France
                \and
                Department of Physics, University of Bath, Claverton Down, BA2 7AY, UK
}

% % The list of authors, and the short list which is used in the headers.
% % If you need two or more lines of authors, add an extra line using \newauthor
% \author[R. S. Beckmann et al.]{R. S. Beckmann$^{1}$\thanks{Email: ricarda.beckmann@iap.fr}, Y. Dubois$^{1}$ et al
% \\
% % List of institutions
% $^{1}$ Institut d'Astrophysique de Paris, CNRS \& Sorbonne Universites, UMR7095, 98bis Boulevard Arago, F-75014, Paris, France \\
% }

% These dates will be filled out by the publisher
\date{Accepted XXX. Received YYY; in original form ZZZ}

% Enter the current year, for the copyright statements etc.

% \abstract{}{}{}{}{} 
% 5 {} token are mandatory
 
  \abstract
   {Extended filamentary H$\alpha$ emission nebulae are a striking feature of nearby galaxy clusters but the formation mechanism of the filaments, and the processes which shape their morphology remain unclear.}
   {We conduct an investigation into the formation, evolution and destruction of dense gas in the center of a simulated, Perseus-like, cluster under the influence of a spin-driven jet. The jet is powered by the supermassive black hole located  in the cluster's brightest cluster galaxy. We particularly study the role played by condensation of dense gas from the diffuse intracluster medium, and the impact of direct uplifting of existing dense gas by the jets, in determining the spatial distribution and kinematics of the dense gas.}
   {We present a hydrodynamical simulation of an idealised Perseus-like cluster using the adaptive mesh refinement code {\sc ramses}. Our simulation includes a supermassive black hole (SMBH) that self-consistently tracks its spin evolution via its  local accretion, and in turn drives a large-scale jet whose direction is based on the black hole's spin evolution. The simulation also includes a live dark matter (DM) halo, a SMBH free to move in the DM potential, star formation and stellar feedback.}
   {We show that the formation and destruction of dense gas is closely linked to the SMBH's feedback cycle, and that its morphology is highly variable throughout the simulation.  While extended filamentary structures readily condense from the hot intra-cluster medium, they are easily shattered into an overly clumpy distribution of gas during their interaction with the jet driven outflows. Condensation occurs predominantly onto infalling gas located 5 - 15 kpc from the center during quiescent phases of the central AGN, when the local ratio of the cooling time to free fall time falls below 20, i.e. when $t_{\rm cool}/t_{\rm ff} < 20$.}
   {We find evidence for both condensation and uplifting of dense gas, but caution that purely hydrodynamical simulations struggle to effectively regulate the cluster cooling cycle and produce overly clumpy distributions of dense gas morphologies, compared to observation.}

   \keywords{Galaxies: clusters: intracluster medium --
                Galaxies: jets --
                Galaxies: clusters: general --
                Methods: numerical --
                Hydrodynamics
               }

   \maketitle

%%%%%%%%%%%%%%%%%%%%%%%%%%%%%%%%%%%%%%%%%%%%%%%%%%

%%%%%%%%%%%%%%%% BODY OF PAPER %%%%%%%%%%%%%%%%%%
\section{Introduction}
One of the most striking features of the nearby Perseus cluster, NGC1275, is the extended filamentary H$\alpha$ emission nebula in its center \citep{Lynds1970,Heckman1989,Crawford1992,Conselice2001,Hatch2007,Fabian2008}. Harbouring up to  $5 \times 10^{10} \rm \ M_\odot$ of cold gas \citep{Salome2006}, this emission nebula has a filamentary morphology, with individual filaments up to 40 kpc long and only 70 pc wide \citep{Conselice2001,Fabian2016}. Within the extended, filamentary H$\alpha$ emission, dense clumps of molecular gas have been observed \citep{Salome2006,Lim2012}, and some filaments show signs of star formation \citep{Fabian2008,Canning2010,Canning2014}. Larger observational samples have shown that the Perseus cluster is not the only object to house such H$\alpha$ emission nebulae, with many massive galaxy clusters showing similar features \citep{Crawford1992,Heckman1989,Mcdonald2010,McDonald2012,Olivares2019} in their center. Where does this gas come from, and what causes its characteristic filamentary morphology?

Finding cold gas in cluster centers is not unexpected. As cooling times in the intra-cluster medium (ICM) of massive galaxy clusters are short, a massive cooling flow of the order of $100 - 1000 \,\rm M_\odot\,yr^{-1}$ is expected to develop in the cluster center \citep{Fabian1994}. However, observed star formation rates in clusters are of the order of only 1-10\% of the naive cooling rate inferred from X-ray observations \citep{McDonald2018}.
Clusters must therefore contain a heating source which prevents overcooling and slows down star formation. Many clusters show evidence for extended jets powered by active galactic nuclei (AGN), which are inflating large cavities in the ICM whose power is sufficient to offset cooling \citep{McNamara2007,Rafferty2006,Fabian2012}. Via the self-regulation cycle, which consists of cold gas feeding the AGN, which in turn powers a jet, which then inflates cavities that heat the ICM, AGN are expected to play a decisive role in determining the cooling and star formation properties of the cluster \citep[see][for a review]{McNamara2007,Fabian2012}.
This picture of self-regulation cycles from AGN jets is getting increasing support from hydrodynamical simulations both in an idealised~\citep{Cattaneo07, Gasparietal11,Li&Bryan14a} and in a cosmological context~\citep{Duboisetal10}.

The cospatiality of the H$\alpha$ emission nebula with the AGN jets and bubbles suggests that the AGN might not only control global cooling properties of the cluster but also be more directly responsible for the morphology of the existing dense gas \citep{Salome2006,Russell2016,Vantyghem2017,Vantyghem2018,McKinley2018,Tremblay2018}. The often complex line-of-sight velocity field of the nebula in Perseus also suggests that this gas is not merely free-falling, or rotationally supported \citep{McDonald2012,Gendron2018}, but most likely interacts with the turbulence injected by the AGN jets and buoyantly rising bubbles \citep{Fabian2003,Hatch2006,Revaz2008}. However, with only line-of-sight velocity information, the three-dimensional velocity pattern of gas is difficult to ascertain. 

Simulations by \citet{McCourt2011} and \citet{Sharma2012} showed that even for a globally thermally stable ICM (required to avoid overly strong cooling flows) dense gas can condense out of the hot ICM via local thermal instabilities for sufficiently low values of $t_{\rm cool}/t_{\rm ff}$. Here, $t_{\rm cool} $ is the local cooling time 
    \begin{equation}
        t_{\rm cool} = \frac{3}{2} \frac{n k_B T}{n_e n_i \Lambda },
    \end{equation}
 where $n_i$, $n_e$ and $n$ are the ion, electron and total number density respectively, $T$ is the temperature and $\Lambda$ the cooling rate. The free fall time is
    \begin{equation}
        t_{\rm ff} = \left (\frac{2r}{g} \right)^{\frac{1}{2}},
    \end{equation}
where $g$ is the local gravitational acceleration and $r$ is the radius from the cluster center. Condensation into multi-phase can take place when locally $t_{\rm cool}/t_{\rm ff}<1$, but it is also observed for larger values of the radial $t_{\rm cool}/t_{\rm ff}$ profile due to the turbulence and inhomogeneities injected by uplifting hot gas from the cluster center via AGN driven feedback processes \citep{Voit2017,Voit2018}. It has been confirmed observationally that molecular gas is observed at the minima of $t_{\rm cool}/t_{\rm ff}$ profiles \citep{Hogan2017,Pulido2018,Olivares2019}, with some of these authors stressing that only $t_{\rm cool}$ determines condensation rates as the growth of linear perturbations is largely independent of the geometry of the gravitational potential \citep{Choudhury2016}.

Simulations have shown that the turbulence injected by AGN feedback can cause the local thermal instabilities predicted by \citet{McCourt2011}, but struggle to reproduce the observed morphologies, with dense gas having either a very clumpy morphology \citep{Li&Bryan14b,Yang&Reynolds16} or settling into a massive central disk  \citep{Gasparietal12,Li&Bryan14a,Li&Bryan14b,Prasad2015}. While the latter has been observed in some clusters, such as in Hydra-A \citep{Hamer2014,Olivares2019}, only a small central disk is observed in Perseus \citep{Nagai2019}. The dense gas morphology therefore seems to sensitively trace the energy balanced in the ICM.

One feature of these clusters is that the cold gas is expected to rain down on the AGN in a cold and chaotic fashion \citep{Gasparietal13,Voit2015,Voit2017}, so the cold gas accreted by the black hole is expected to lack coherent angular momentum, which in turn could lead to a reorientation of the black hole spin axis over time. In this paper, we investigate the impact of this chaotic dense accretion on the formation of further gas by explicitly tracing the spin of the black hole, and using this black hole spin axis as the axis for the AGN driven jet \citep{Duboisetal14bhspin}. In contrast to existing simulations, which rely on a fixed jet axis with pre-defined precession within a narrow jet cone \citep{Li&Bryan14a,Yang&Reynolds16,Ruszkowskietal17,Lietal17,Prasad2018,Martizzietal19,Wangetal19}, 
the spin driven approach used here is able to inject turbulence over a larger volume of the cluster center, and to respond dynamically to the evolving dense gas morphology throughout the simulation. 

In this paper, we will investigate the formation and time evolution of dense gas structures in a Perseus-like cluster under the influence of a spin-driven jet, with a particular focus on clump dynamics. The simulations are introduced in section \ref{sec:setup}. A general overview of results is given in section \ref{sec:cluster_evolution}, the jet evolution is studied in Section \ref{sec:jet} and the clump properties are investigated in Section \ref{sec:clump_properties}. A detailed look at the role of uplifting in clump properties and dynamics is given in Section \ref{sec:uplifting}, and the impact of condensation is studied in Section \ref{sec:condensation}. A discussion of results can be found in section \ref{sec:discussion}, and conclusions are summarised in \ref{sec:conclusions}.

\section{Simulation setup}
\label{sec:setup}

% Calculations for values to be found in http://localhost:8888/notebooks/2018_10_31_RESULTS_compute_paper_quantities.ipynb

This paper presents a set of hydrodynamical simulations of isolated galaxy clusters, produced with the adaptive mesh refinement code \textsc{ramses}  \citep{Teyssier02}. 

\subsection{Technical details and refinement}
\label{sec:refinement}
For the simulations presented here, the Euler equations are solved with the second order MUSCL-Hancock scheme, which computes Godunov fluxes using an approximate HLLC Riemann solver and a MinMod total variation diminishing scheme to reconstruct the interpolated variables.
The Courant factor is set to a value of $0.8$.

The simulation is performed in box of size $8\, \rm Mpc$ with a root grid of $64^3$, and then adaptively refined to a maximum resolution of $120 \rm \ pc$. Refinement proceeds according to several criteria. We use a quasi-Lagrangian criterion: when a cell contains a mass greater than $3.5\times 10^{9} \,\rm M_\odot$, it is refined ( and it is derefined if it contains less than 0.125 this). We also use a Jeans-based criterion: a cell is refined until the local Jeans length is $> 4$ times the cell size. To refine regions of interest to this work, we also employ two additional refinement criteria. First, the cell containing the BH is forced to be refined at the maximum resolution. Second, a passive scalar variable is injected by the BH jet with a mass density $\rho_{\rm scalar}$ equals to that of the gas $\rho_{\rm gas}$, which is advected with the gas and marks regions affected by BH feedback. The scalar decays exponentially, with a decay time of $t_{\rm jet} = 10 \rm \ Myr $ to ensure that the scalar traces only recent AGN feedback events. After testing different decay times, we can confirm that the results do not sensitively depend on this value.
To resolve the AGN bubbles, cells are allowed to be further refined when the scalar fraction exceeds $\rho_{\rm scalar}/\rho_{\rm gas}>10^{-4}$, equivalent to 92 Myr since the last feedback event, and its relative variation from one cell to another exceeds $10^{-2}$. The latter two refinement criteria ensure that the regions affected by AGN feedback, including the hot, low density bubbles which would de-refine under a purely Lagrangian refinement scheme, remain maximally refined over a reasonable duration of the jet propagation and mixing with hot ICM.

\subsection{Initial conditions}
The initial conditions for dark matter (DM) and gas consist of a cored Navarro-Frenk-White profile:
\begin{equation}
\rho_{\rm DM}=(1-f_{\rm gas})\rho_{\rm s}\frac{r_{\rm s}^3}{(r+r_{\rm core})(r+r_{\rm s})^2} \, , 
\label{eq:NFW}
\end{equation}
where $r_{\rm s}=r_{200}/c$ is the scale radius, $r_{\rm core}=20 \rm \ kpc$ is the core radius, $\rho_{\rm s}=\rho_{\rm c}\delta_{200}$ is the density scaling of the profile, with $\rho_{c}$ the critical density of the Universe. The rescaling factor $\delta_{200}=\frac{200\Omega_{m}}{3}\frac{c^3}{f(c)}$, where $f(c)=\ln(1+c)-c/(1+c)$, to the radius at which the average density of the profile is 200 times the mean density of the Universe. $f_{\rm gas}$ is the gas fraction of the halo, here taken to be 15~\%.
The halo has a concentration parameter $c=6.8$, and a virial velocity $v_{200} = 1250\, \rm km\, s^{-1}$.

DM particles have a mass resolution of $3.7 \times  10^8 \rm \ M_{\odot}$ and are distributed using the {\sc dice} code~\citep{Perretetal14}.
The profile is truncated at a radius 2.2 Mpc, for a total DM halo mass of $3.4 \times 10^{14} \rm \ M_\odot$. DM particle are live and able to move under gravity, allowing the DM potential to respond to the evolution of the cluster core throughout the simulation.

Gas is initiated in hydrostatic equilibrium assuming a gas fraction of $15~\%$ distributed according to the profile of the DM (see Eq. \ref{eq:NFW}), and then allowed to cool. As part of the initial conditions, turbulence is injected into the gas with a velocity dispersion of $15 \rm \ km\, s^{-1}$, but no rotation is added to the halo. This small initial velocity dispersion in the hot gas serves to break the symmetry of the initial conditions. Metallicity is initially set to $0.3 \rm \ Z_\odot$ throughout, and the BH sinkparticle is placed in the centre of the halo. No stars are added as part of the initial conditions. In order to avoid edge effects, the halo is placed in a sufficiently large box (8 Mpc on a side), and initiated with a gas density of $9.8 \times 10^{-8} \rm \ cm^{-3}$  outside of the truncation radius of the halo. 

\subsection{Cooling}
The metal-dependent cooling of the gas is followed using the tabulated values of~\cite{Sutherland&Dopita93} down to $10^4 \, \rm K$.
The cooling function is extended below $10^4\, \rm K$ with the fitting functions from~\cite{Rosen&Bregman95}.
Solar abundance ratios of the elements is assumed throughout independent of the overall metallicity.

\subsection{Star formation and stellar feedback}
Star formation proceeds according to a combined density and temperature criterion, with star formation permitted in cells with hydrogen number density of $n_{\rm H} > 1 \rm \, H\,  cm^{-3} $ and temperature ${T < 10^4 }$ K. The mass resolution of stars is $n_{\rm H}m_{\rm p}\Delta x^3/X_{\rm H}= 5.6 \times 10^{4} \rm \ M_\odot$, where $X_{H}=0.74$ is the fractional abundance of hydrogen.
The star formation rate density proceeds according to a Schmidt law $\dot \rho_*=\epsilon_*\rho/t_{\rm ff}$, where $\rho$ is the gas density, $t_{\rm ff}$ is the gas free-fall time, and $\epsilon_*=0.1$ is the constant efficiency of star formation.

Stellar feedback is included in the form of type II supernovae only. We use the energy-momentum model of \citet{Kimmetal15} with each stellar particle releasing  an energy of $e_{\rm *,SN} =  m_* \eta_{\rm SN}10^{50}\, \rm erg\, M_\odot^{-1}$ at once after $10\, \rm Myr$, where $\eta_{\rm SN}=0.2$ corresponding to the mass fraction of the initial mass function for stars ending up their life as type II supernovae, and $m_*$ is the stellar particle mass. These explosions also enrich the gas with metals with a constant yield of 0.1. Metals are treated as a single species and are advected as a passive scalar.

\subsection{Black hole accretion and feedback}
\label{sec:AGN_method}
AGN feedback from the central BH is followed using the model from~\cite{Duboisetal10} with several modifications that include the self-consistent evolution of the BH spin~\citep{Duboisetal14bhspin} and the spin-dependent feedback efficiency~(Dubois et al., in prep.). 

A BH ``sink'' particle is placed at the centre of the halo as part of the initial conditions, with a mass of $3.4 \times 10^8 \rm \ M_\odot$. The BH is then free to move across the grid throughout the simulation.
To compensate for unresolved dynamical friction from the stars within the host galaxy, an analytic drag force is applied to the sink particle according to \citet{Pfisteretal19}. We do not model the equally unresolved gas drag explicitly as the difficulty in measuring the relative velocity between the sink and the turbulent ISM introduces too many numerical artifacts in the black holes trajectory \citep[see][]{Beckmann2018b}. A particular worry is the black hole getting attached to its own feedback and ejected from the central galaxy, which we avoid here by not using a sub-grid prescription for the gas drag.

The BH accretes according to the Bondi-Hoyle-Lyttleton accretion rate 
\begin{equation}
\dot{M}_{\rm BHL} = \frac{G^2 M_{BH}^2 \bar{ \rho}}{(\bar{c_s}^2+\bar{v}^2)^{3/2}}\, ,
\end{equation}
where $\bar{\rho}$, $\bar{c_s}$ and $\bar{v}$ are the mass weighted local average density, sound speed and relative velocity between the gas and the BH. All quantities are measured within a sphere with radius $4 \Delta x_{\rm min}$ centered on the instantaneous position of the BH, with the BH free to move across the grid. $\Delta x_{\rm min}$ is the size of the smallest resolution element of the simulation. Accretion is not limited to the Eddington accretion rate. 

The AGN feedback is modelled with jets following the injection method from~\cite{Duboisetal10}. 
At each feedback event, feedback energy 
\begin{equation}
    \dot{E}_{\rm feed} = \eta_{\rm MAD} \dot{M}_{\rm BHL}c^2
    \label{eq:mad}
\end{equation}  (where $c$ is the speed of light) is injected as kinetic energy within all cells contained in a cylinder of radius 0.4 kpc and height 0.8 kpc. The cylinders is aligned with the BH spin axis.
The efficiency $\eta_{\rm MAD}$ is a spin-dependent efficiency obtained from magnetically arrested disc (MAD) simulations from~\cite{McKinneyetal12}, which has a minimum at a spin of 0, and a maximum at a spin of 1.
The BH spin-up rate is taken from the same simulations.
The AGN jet is always taken to be aligned with the BH spin axis, and the conditions for BH-disc alignment in misaligned grid-scale gas angular momentum (with that of the BH spin) 
is obtained by Lense-Thirring considerations~\citep[see][for details]{Duboisetal14bhspin}.
As the spin-axis changes self-consistently throughout the simulation, we do not need to add any explicit precession to the jet, as it naturally arises from the chaotic nature of the cold gas accretion onto the BH~\citep{Gasparietal13}, which regularly changes the BH spin direction over time (see Section \ref{sec:jet}).

As mentioned in Section \ref{sec:refinement}, a passive scalar of density $\rho_{\rm scalar} =  \rho_{\rm gas}$ is injected within the feedback cylinder at each feedback event, where $\rho_{\rm gas}$ is the gas density. This scalar then decays exponentially with a decay time of 10 Myr, allowing cells recently affected by the AGN jet to be identified. Therefore, with the AGN passive scalar quantity, one can define an age for the gas that has been impacted by the AGN, with ${t_{\rm AGN}=-10\ln{(Y_{\rm scalar})}\, \rm \ Myr}$, where $Y_{\rm scalar}=\rho_{\rm scalar}/\rho_{\rm gas}$.

\subsection{Tracer particles}
\label{sec:tracers}
To follow the dynamical history of gas in the simulation we employ Monte-Carlo based tracer particles from~\cite{Cadiouetal19}. These tracer particles are a significant improvement over classical ``velocity''-based tracer particles, in particular in regions with strongly converging flows such as cold gas condensation and gas collapsing under self-gravity.
We set up $2 \times 10^8$ tracer particles, with each particle tracing a gas mass of $4\times 10^5 \rm \ M_\odot$.
They are initially distributed according to the gas density profile in the initial conditions, out to a radius of $200 \rm \ kpc$.

\section{Results}

\subsection{Cluster evolution}
\label{sec:cluster_evolution}

\begin{figure*}
% Made using 2019_01_16_multiplot_per_output.py
    \centering
    \includegraphics[width=0.9\textwidth]{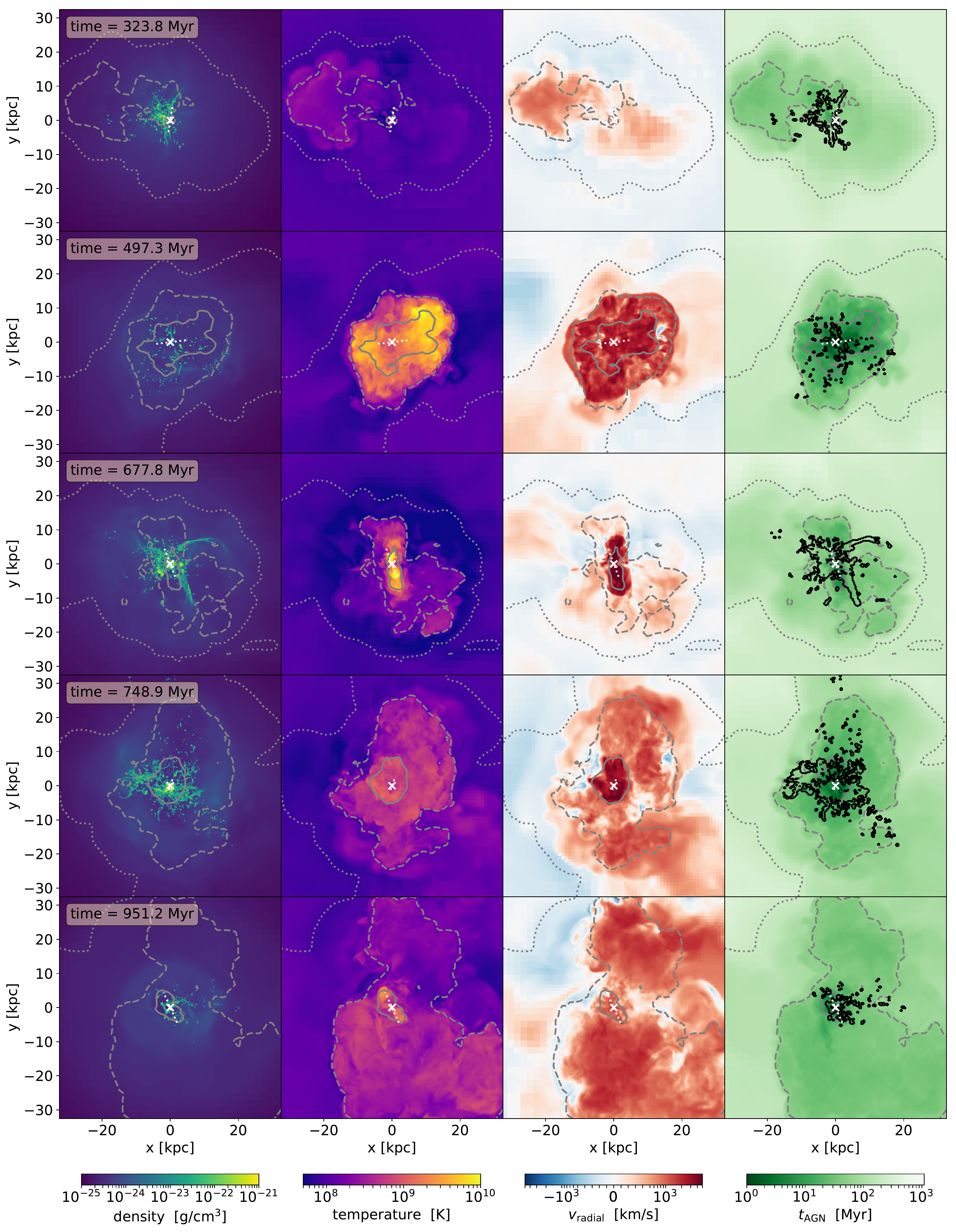}
    \caption{Projections of (from left to right) density, temperature, radial velocity and the time since a cell has been affected by AGN feedback, $t_{\rm AGN}$, at five different points in time. Radial velocity and temperature are weighted by $t_{\rm AGN}$. Radial velocity is measured in 3D space with the SMBH at the origin, and negative velocities are inflowing.  Contours are based on the plot of $t_{\rm AGN}$, and are drawn at 10 (solid), 50 (dashed) and 200 (dotted) Myr. The location of the SMBH is marked by a white cross, and black contours in the right hand column denote the outline of dense gas structures. The white dotted line lies along the instantaneous jet axis, which is plotted to be exactly 10 kpc long in 3D space. The shorter it appears, the more it is aligned with the line of sight of the image (here taken to be the z-axis of the box at all times).
    }
    \label{fig:pretty_picture}
\end{figure*}

As can be seen in Fig.~\ref{fig:pretty_picture}, which shows the gas density, temperature, AGN age (see section~\ref{sec:AGN_method}) and the gas radial velocity at various times, the gas in the cluster develops a multi-phase structure with a complex morphology that evolves significantly over the course of the simulation. 

The hot gas in the intra-cluster medium, which has temperatures in the range $0.09 - 1131$ keV ($10^6 - 1.3\times10^{10}$ K), cools down and condensates into dense clumps and filaments within the central 50 kpc of the cluster, with an average temperature of the dense gas of $4.0 \times 10^{-4}$ keV ($4.6 \times 10^3$ K). This dense gas falls towards the center where it feeds the central BH and thereby triggers the AGN jet, which, in return, interacts with existing dense gas and stirs turbulence into the hot gas, generating hot outflows with  outflow velocities up to ${3.5 \times 10^4 \rm \ km\, s^{-1}}$. As the radio jet is oriented along the SMBH spin axis, which in turn is updated according to the chaotic cold accretion onto the central BH~\citep{Gasparietal13,Voit2017}, the jet continuously re-orients throughout the simulation (see Section \ref{sec:jet} for details). As a result, the shapes of the jet relics indicated by the ``AGN age'' also change significantly over time.

\begin{figure}
%made using 2018_10_31_RESULTS_cluster_property_time_evolution.ipynb on horizon
    \centering
    \includegraphics[width=\columnwidth]{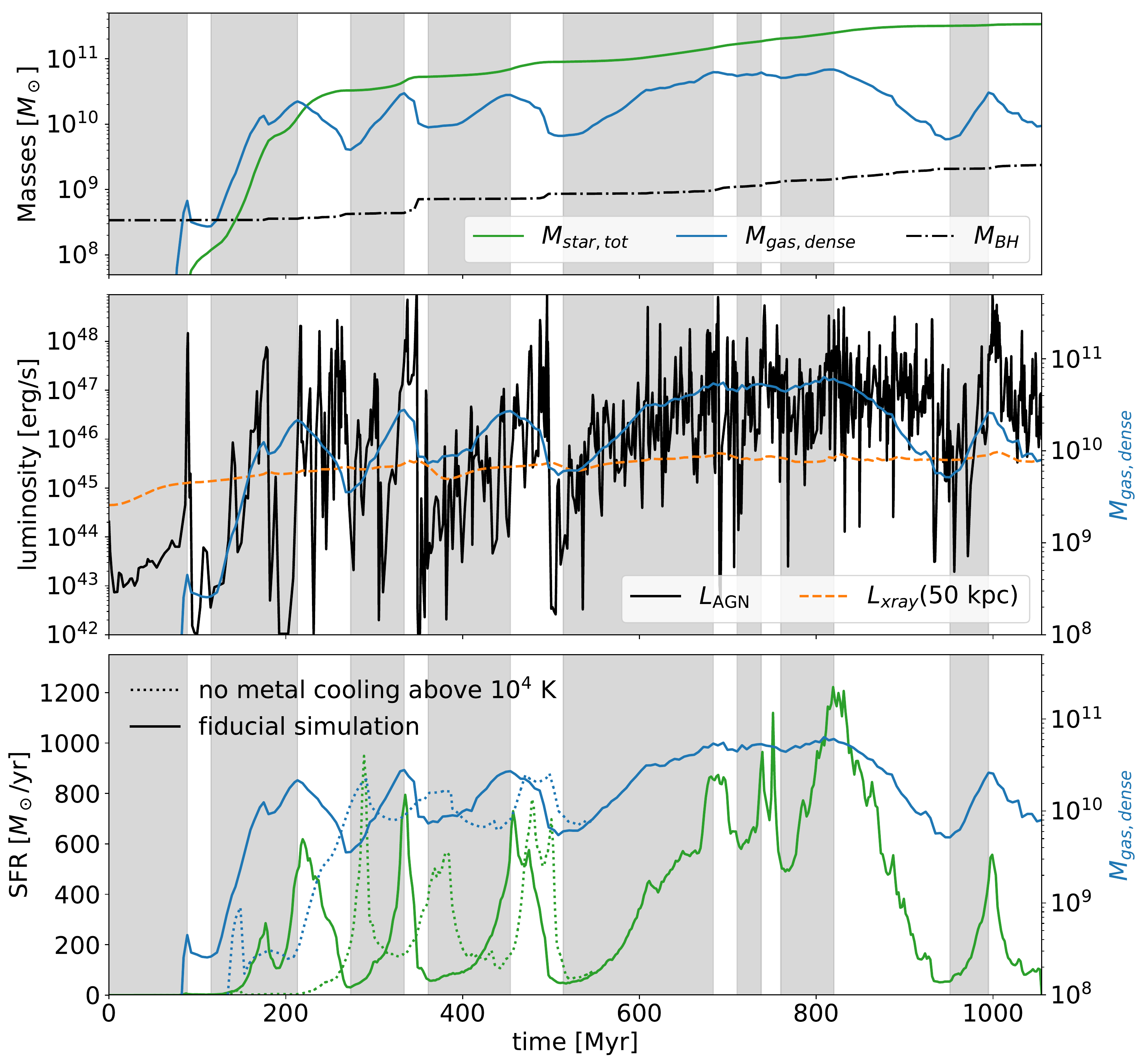}
    \caption{{\em Top panel:} Time evolution of cluster properties including stellar mass $M_{\rm star}$, BH mass  $M_{\rm BH}$ and gas mass  $M_{\rm gas}$. {\em Middle panel:}  AGN luminosity, X-ray luminosity of hot gas within 50 kpc of the cluster center, and the dense mass again for comparison. {\em Bottom panel:} SFR, as well as the dense gas mass again for comparison, for both the fiducial simulation and for a companion simulation without metal cooling for gas with $T>10^4$ K (see text for details). Dense gas is defined to be gas with a temperature at or below $T_{\rm dense}= 10^6$ K, hot, diffuse gas with a temperature above that. White and grey background colours show the heating and cooling dominated regimes of the fiducial simulation.}
    \label{fig:masses_timeseries}
\end{figure}

More quantitatively, Fig. \ref{fig:masses_timeseries} shows that gas begins to cool after approximately 100 Myr, equivalent to the initial cooling time of gas in the cluster center as set by the initial conditions. Dense gas, for the remainder of the analysis, is defined to be gas with a maximum temperature of $10^6$ K. By 139 Myr, the dense gas mass first exceeds $10 ^ 9 \rm \ M_{\odot}$, and the cluster enters a cyclic behaviour where dense gas repeatedly builds up to a total mass in excess of $2 \times 10^{10} \rm \ M_\odot$ before being reduced to closer to $2 \times 10^{9} \rm \ M_\odot$.

We have split the evolution of the cluster into two regimes using the total dense gas mass. A cooling dominated regime, when the total dense gas mass of the cluster increases (marked with a grey background in Fig. \ref{fig:masses_timeseries}), and a heating dominated regime, when the total dense gas mass of the cluster decreases. The regime of the cluster is evaluated using the smoothed derivative of the mass of dense gas $\rm M_{gas,dense}$. The total dense gas mass in the cluster can be reduced in a number of different ways: dense gas can be consumed in star formation, accreted onto the BH or destroyed via hot winds or shocks driven by AGN feedback.

AGN activity (see second panel of Fig. \ref{fig:masses_timeseries}) is highest during the heating-dominated phase, with maxima in dense gas followed by maxima in AGN activity within 50 Myr or less. These peaks in AGN activity destroy dense gas in the cluster, causing the AGN to enter a low feedback state until the dense gas mass has had time to build up again. Only a small fraction of the gas is directly accreted by the SMBH, as can be seen by the fact that the mass increase of the SMBH mass in the top panel of Fig.~\ref{fig:masses_timeseries} is much smaller than the decrease in dense gas mass over the equivalent period of time. 

As can be seen in the bottom panel of Fig. \ref{fig:masses_timeseries}, the star formation rate varies strongly over time, following the general trends set by the total dense gas mass in the cluster. There are clear bursts of star formation in the cooling dominated interval. This suggests that a significant amount of the dense gas is directly consumed by star formation. At peaks of up to $1000 \rm\ M_\odot yr^{-1}$, the star formation rate of our simulated cluster is extremely high in comparison to observations, which for equivalent mass clusters report star formation rates in the range $ 1 - 100 \rm\ M_\odot\,yr^{-1}$ \citep{Odea2008}. The dense gas mass, by contrast, falls close to the $10^{10}-10^{11} \rm \ M_{\odot}$ observed in Perseus \citep{Bridges1998,Salome2006,Mittal2015}. The SFR might be so elevated in comparison to observations because gas is cooling too efficiently to start with, or because gas is being converted too efficiently into stars once cooled. The latter is discussed further in Sec. \ref{sec:sf_efficiency}.

If gas in the cluster is cooling too efficiently, too much gas is transitioning from the hot, diffuse phase to the dense phase. The X-ray luminosity of the central 50 kpc of the simulated cluster are in the range of $1.2-5.3 \times 10^{45} \, \rm erg\, s^{-1}$ , with the observed values for Perseus of $1.26 \times 10^{45} \rm \, s^{-1}$  \citep{Ebeling1996} at the lower end of that range. While the initial conditions were chosen to reproduce observed profiles, the emitted X-ray luminosity increases due to the gas cooling in the cluster center.

 One limitation of the simulation presented here is the lack of cosmological context, which means that heating due to turbulence injected by large-scale phenomena, such as galaxy mergers or anisotropic accretion, is absent. In addition, being purely hydrodynamical, the simulation disregards effects such as magnetic fields and other non-thermal energy sources such as cosmic rays, which could heat the gas and provide an extra pressure support against collapse on small scales. 

Currently, we rely on equilibrium cooling with an initially uniform metallicity of 0.3 $Z_\odot$ everywhere, based on observations of the metallicity in the outskirts of Perseus by \cite{Werner2013}.
By 1 Gyr, the volume weighted metallicity in the central 50 kpc of the hot ICM has risen to 0.36 $Z_\odot$ due to stellar feedback. While this is higher than the initial value, it still falls below the value of 0.6 observed by \cite{Schmidt2002}. 
One possibility is that equilibrium cooling assumed here over-estimates the contribution of metal cooling at high temperatures. X-rays emitted by the AGN could dissociate metals in high temperature gas, reducing their contribution to cooling \citep{Dubois2011,Agertz2013}.

If radiative transfer and non-equilibrium processes were included, the hard X-rays emitted by the AGN would be able to photo-ionize some important metal coolants further so that their contribution to cooling is reduced~\citep[e.g.][]{Gnedinetal12,Segersetal17}. As metal line cooling is the dominant cooling channel for gas between $10^ 5 - 10^7$ K, shutting down metal cooling would hamper the transition of gas from the hot, diffuse phase to the dense phase. To test this hypothesis, we ran a simulation using a cooling function in which the metal cooling function is modified by a kernel 
\begin{equation}
f_{\rm cool}= \exp \left[- \left(\frac{\rm T}{10^4\, \rm K}\right)^{10} \right],
\end{equation}
which effectively shuts off metal cooling for gas with temperatures above $\rm T > 10^4 \rm \ K$. As can be seen in the bottom panel of Fig. \ref{fig:masses_timeseries}, while the initial cooling is delayed in comparison to the fiducial simulation, SFRs remain high even with truncated metal line cooling and the evolution of dense gas is qualitative indistinguishable between the two simulations. We therefore conclude that metal line cooling is not the root cause of the over-cooling reported here. It is more likely that the over-cooling occurs due the absence of non-thermal energies from cosmic rays, which are expected to be able to offset as much as 60~\% of the thermal cooling in a cluster environment \citep{Pfrommer2013,Jacob2017a,Jacob2017b, Ruszkowskietal17}, while only contributing on the percent level to the overall pressure \citep{Reimer2004,Brown2011}. Due to the large reservoir of heat in cluster outskirts, thermal conductivity in massive clusters can also be an efficient process to bring balance back to the hot cooling gas in the center of clusters~\citep{Narayan01, Ruszkowski11,Yang&Reynolds16Cond, Kannanetal17}.
These avenue of investigation will be explored in future work.

\subsection{Jet evolution and turbulence in the cluster}
\label{sec:jet}
\begin{figure}
    \centering
    \includegraphics[width=0.9\columnwidth]{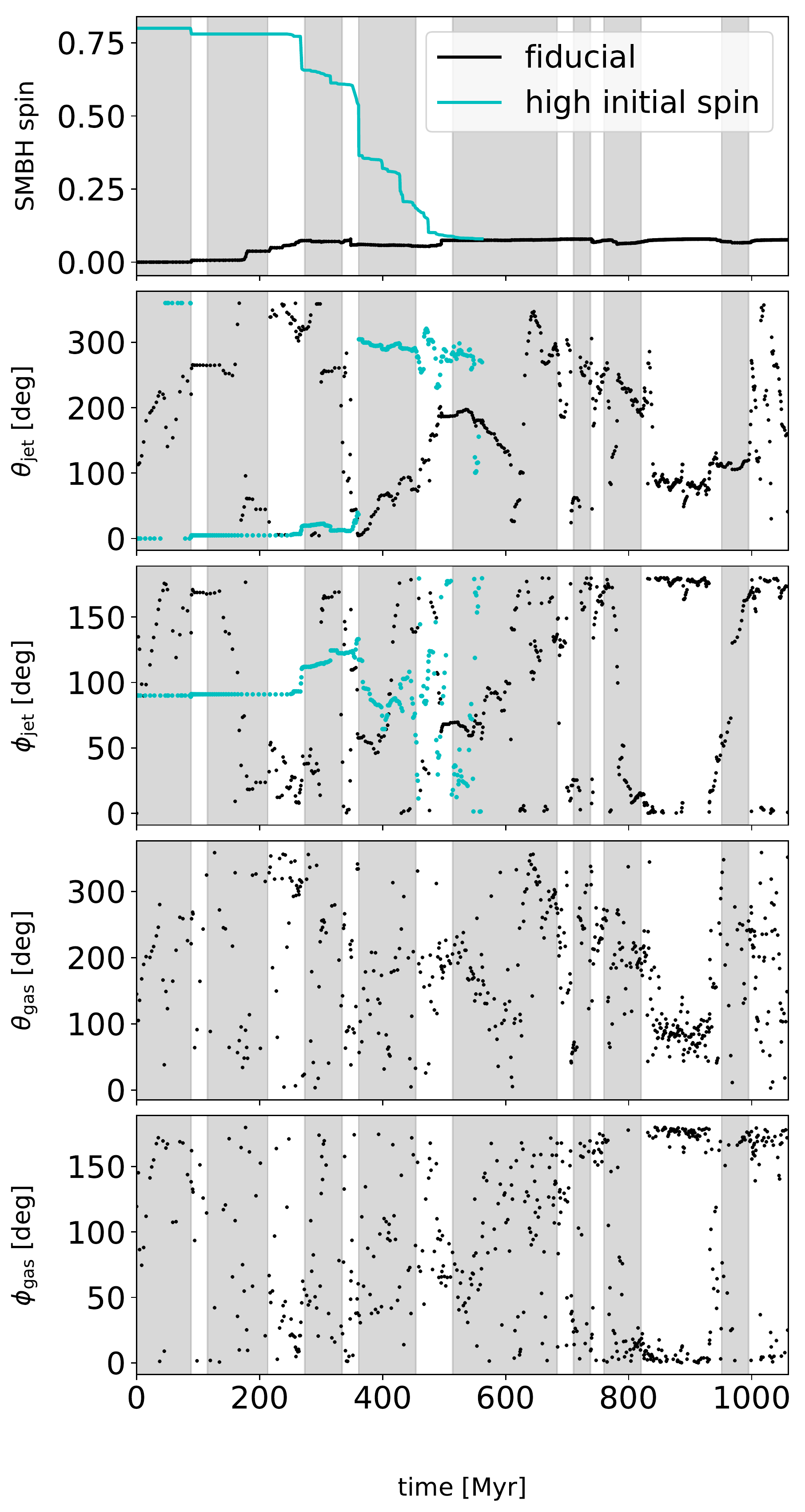}
    \caption{Spin evolution of the SMBH, showing the spin magnitude (top panel), and the two angles defining the SMBH axis (second + third panel) and the angular momentum of the accreted gas at that particular timestep (bottom two panels). The angles are measured in the box frame, and are defined to be the same as in polar coordinates, where $\theta_{\rm jet}$ is measured in the x-y plane of the box (shown in Fig. \ref{fig:pretty_picture}) and $\phi_{\rm jet}$ is the angle with the z-axis (the line of sight in Fig. \ref{fig:pretty_picture}). Angles are measured in the range  $0 \leq \theta < 360^\circ$ and $0 \leq \phi < 180^\circ$. Discontinuous jumps from just below the upper end of the range, to just above the lower end of the range, or vice versa, are due to the cyclic nature of the coordinate system. The top three panels show both the fiducial simulation, and a second, identical simulation initiated with a higher spin value. White and grey background colours show the heating and cooling dominated regimes of the fiducial simulation.}
    \label{fig:spin}
\end{figure}

\begin{figure}
    \centering
    \includegraphics[width=\columnwidth]{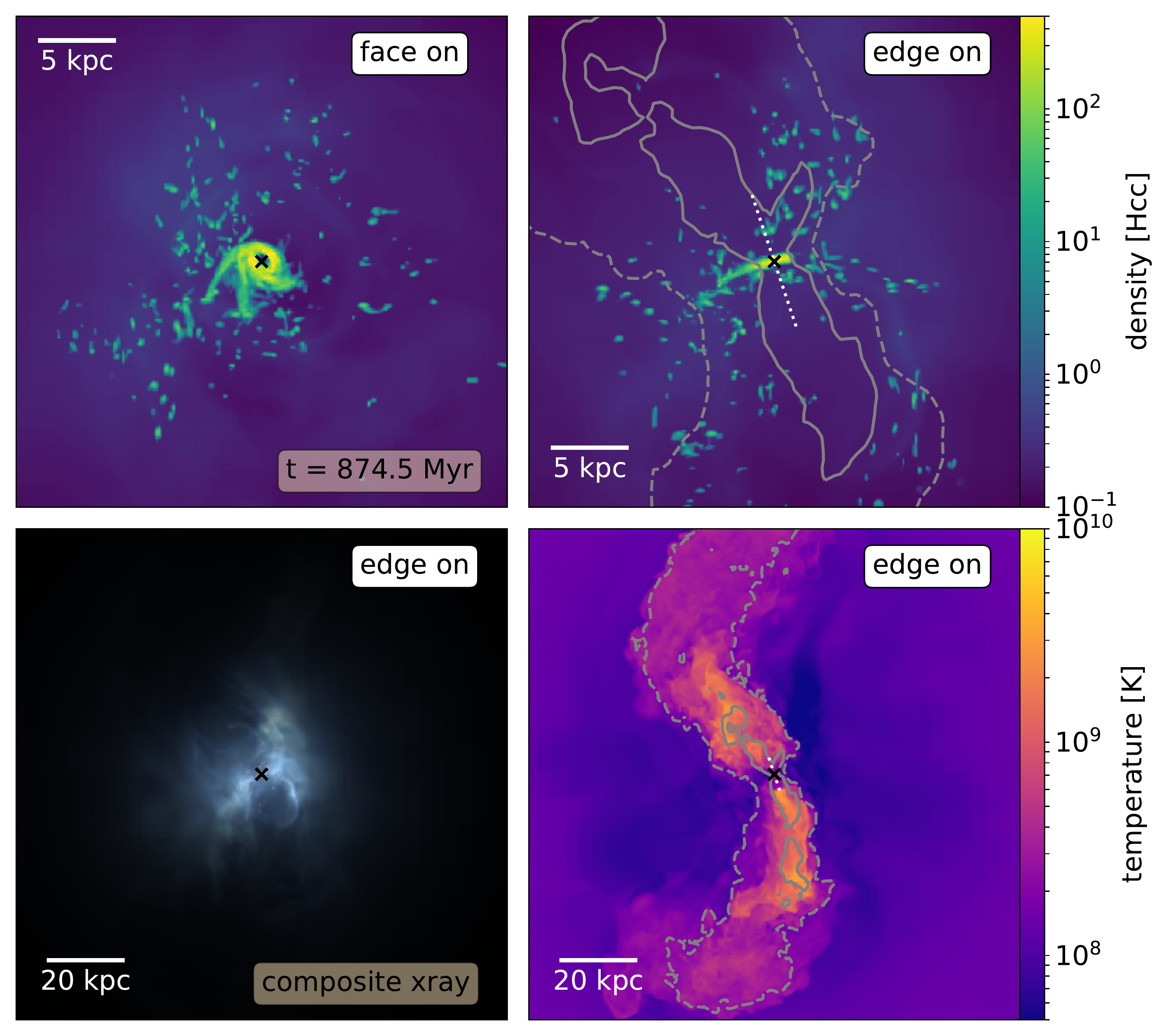}
    \caption{Projection plots at $t=874.5$ Myr, showing the central gas disc in the cluster: Top row - density projections of the cluster center along two different lines of sight, Bottom left: composite x-ray image, using the same x-ray bins as in Figure \ref{fig:xray}, Bottom right: jetscalar weighted temperature projection. Contours mark $t_{\rm AGN} = 10$ and 50 Myr (solid, dashed). The SMBH location is marked by a black cross, and the jet direction is shown by a white dotted line in the right hand panels only.}
    \label{fig:gas_disc}
\end{figure}

One important difference between the work presented here, and previous works on the subject \citep{Li&Bryan14a,Li&Bryan14b,Yang&Reynolds16,Ruszkowskietal17,Lietal17,Cieloetal18,Martizzietal19,Wangetal19} is that our jet axis is not fixed throughout the simulation, nor do we add explicit precession. Instead, the spin evolution of the BH not only determines the feedback energy but also, crucially, the direction of the jet, as the jet axis is taken to be aligned with the BH spin axis, and the BH spin axis is continuously updated according to the angular momentum of accreted gas. 

Fig. \ref{fig:spin} shows that the direction of the jet explores the full parameter space of the simulation, repeatedly traversing the full range of both polar and azimuthal angles ($0 \leq \theta_{\rm jet} < 360^\circ $ and $0 \leq \phi_{\rm jet} < 180^\circ$). This is a consequence of the chaotic angular momentum accreted by the SMBH. As can be seen in the bottom two panels of Fig \ref{fig:spin}, the angular momentum of the accreted gas varies extremely rapidly, both in $\theta$ and in $\phi$, as clumps rain down on the BH from all directions. As the BH spin evolution is a continuous measure, it varies more slowly than the angular momentum of the accreted gas. The only time both the gas angular momentum and the BH spin direction settle occurs in the period $t= 820 - 950$ Myr, when both the BH spin and the angular momentum have $\theta \approx 90^\circ$ and $\phi$ close to zero (the apparent large gap in $\phi$ at this time is a feature of the coordinate system chosen. $\phi=2^\circ$ to $\phi=178^\circ$ only represents a rotation of $4^\circ$, as both $0^\circ$ and $180^\circ$ are aligned with the z-axis of the box). At this time a rotating central gas disc forms around the SMBH, as can be seen in Fig. \ref{fig:gas_disc},which drives jet bubbles out to more than 70 kpc from the cluster center. Dense clumps continue to exist at larger radii, but are preferentially found outside the region recently affected by AGN feedback (see solid grey contours on the image). 

\begin{figure*}
    \centering
    \begin{tabular}{ccc}
        \includegraphics[width=0.3\textwidth]{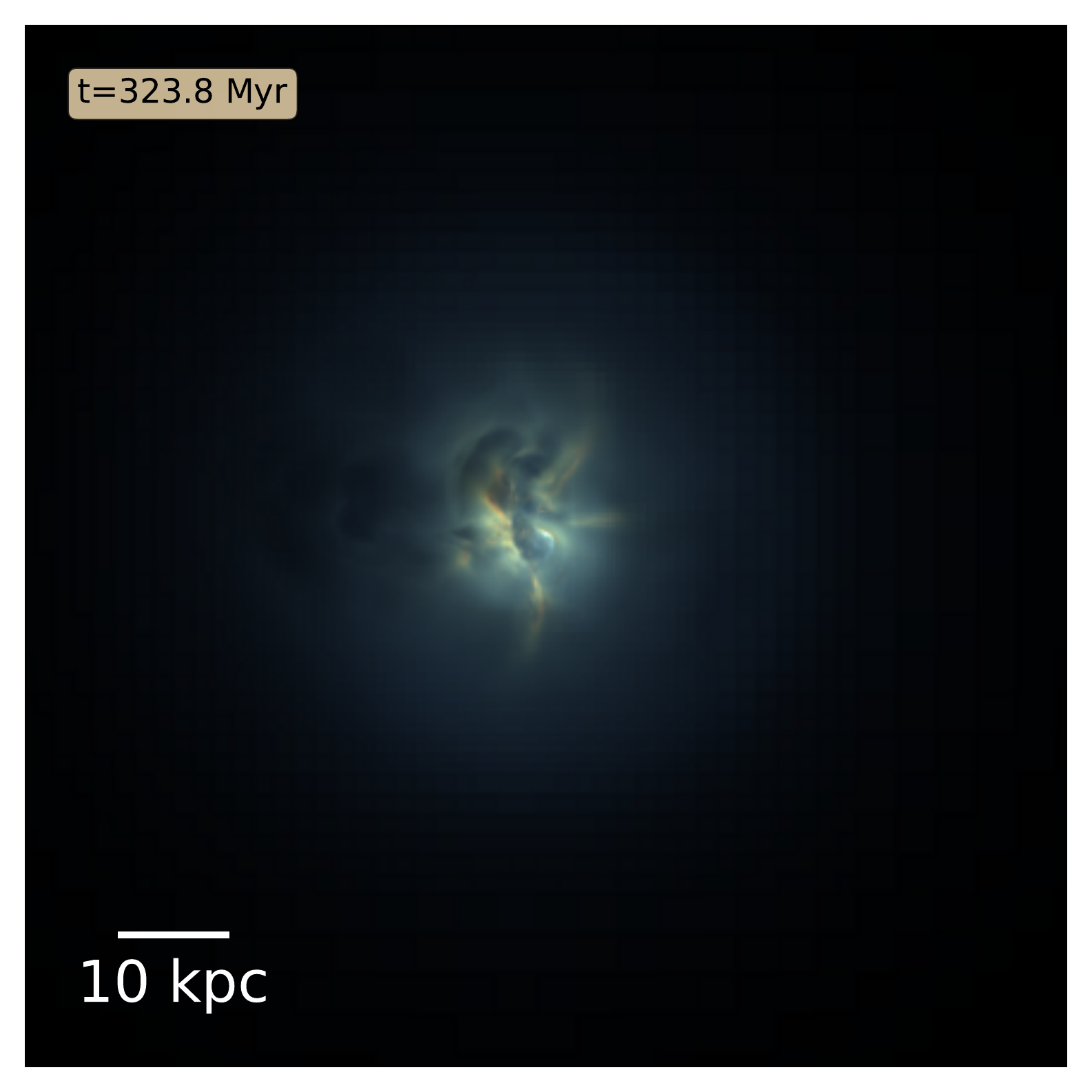}& 
        \includegraphics[width=0.3\textwidth]{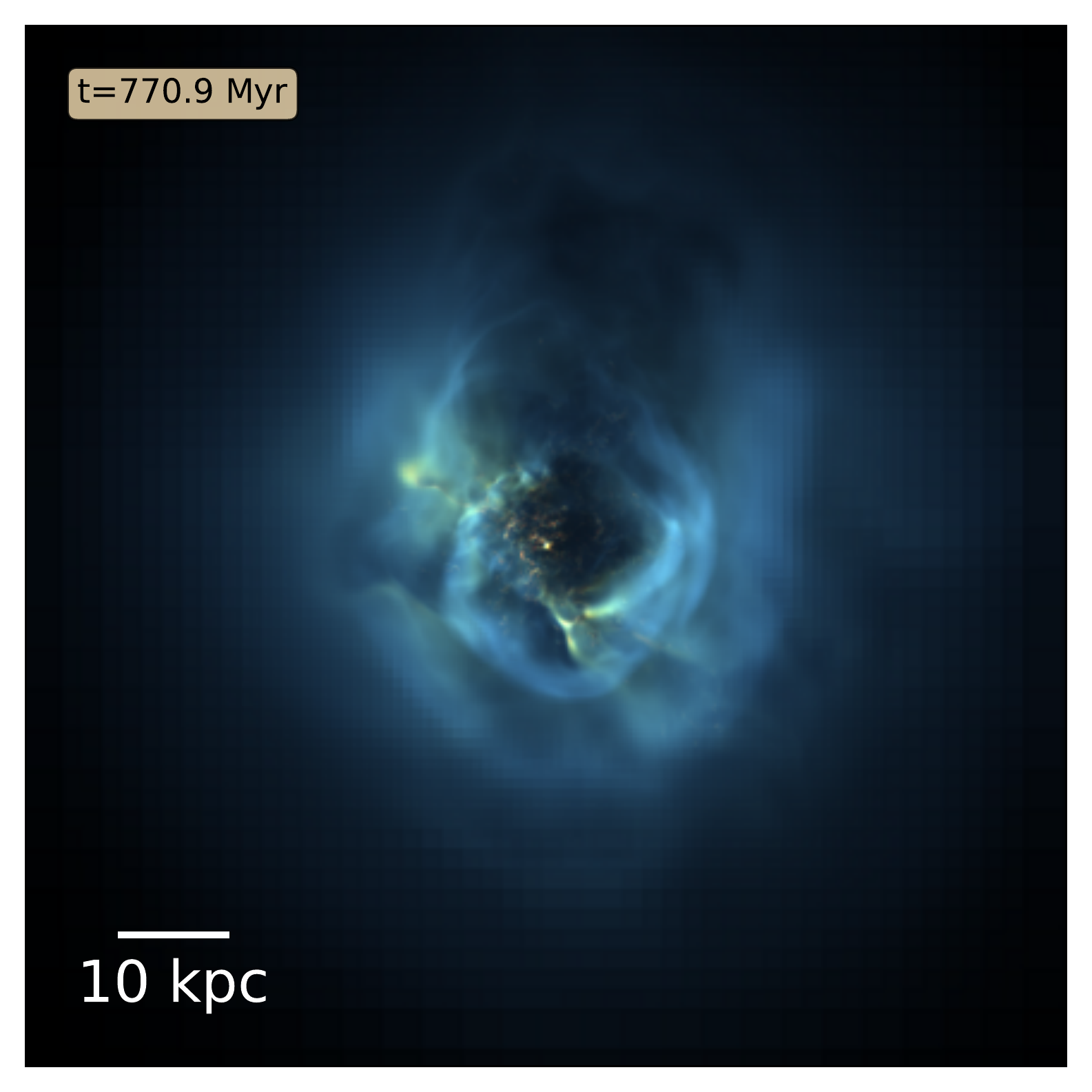}&
        \includegraphics[width=0.3\textwidth]{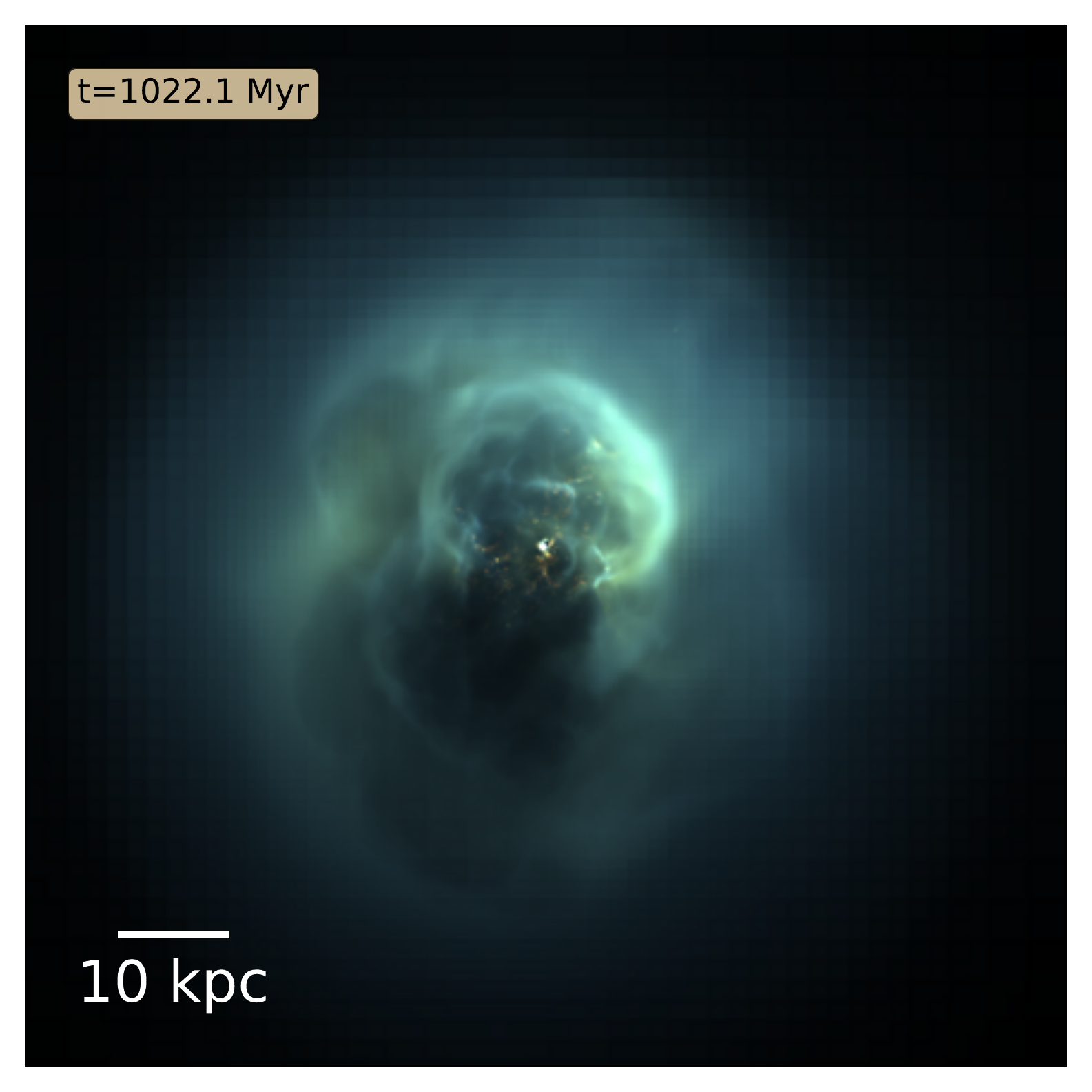}\\
    \end{tabular}

    \caption{Synthetic composite X-ray images of the cluster center, with 0.3-1.2 keV in red, 1.2-2 keV in green and 2-7 keV in blue, to match the image of Perseus in \citet{Fabian2005}, towards the beginning, middle and end of the simulation. Each channel is scaled to highlight fainter features. Each image is 100 kpc across.}
    \label{fig:xray}
\end{figure*}

Our jets self-consistently produce a three-dimensional distribution of fat feedback bubbles seen in Fig. \ref{fig:xray}, without the need for adding an ad hoc precession or reorientation of the jet~\citep[as done in e.g.][]{Li&Bryan14a,Li&Bryan14b,Yang&Reynolds16,Ruszkowskietal17,Lietal17,Cieloetal18,Martizzietal19, Wangetal19}. Firstly, the jet reorientation due to spin helps to self-regulate the cooling flow in clusters~\citep{Cieloetal18} by more uniformly redistributing the energy in the hot gas as long as the reorientation is moderate (i.e. not too close to mimicking isotropic energy input, see~\citealp{Gasparietal12}). Secondly, the reorienting jet has important consequences for the distribution of turbulence in the cluster center, as over time a much larger volume is directly affected by the AGN jet. However, the bubbles shown here are less round and more broken up than observed X-ray cavities in clusters. This is due to the fact that in the absence of viscosity and magnetic fields, strong Rayleigh-Taylor and Kelvin-Helmholtz instabilities at the bubble surface break up bubbles prematurely and shorten their overall lifetime \citep{Ogiya18}.

While we explicitly track the spin evolution of the BH, as described in Section \ref{sec:AGN_method}, the magnitude of the BH spin remains small throughout, as can be seen in Fig. \ref{fig:spin}, with a maximum spin parameter of 0.08. This is partially a consequence of the model chosen, as the MAD jet model always preferentially reduces the spin of the BH. This low spin value in turn has consequences for the jet direction, as the jet axis is aligned with the BH spin axis. Due to the low spin value of the BH, the chaotic angular momentum of accreted gas (see bottom two panels of Fig. \ref{fig:spin}), driven by the chaotic infall of the clumps, is able to significantly realign the spin axis throughout the simulation. 

As can be seen in Eq. \ref{eq:mad}, the feedback energy of the BH is determined by the feedback efficiency $\eta_{\rm MAD}$, which in turn is determined by the BH spin. Due to the consistently low spin-values, the simulation presented here has an average luminosity-weighted feedback efficiency of only 0.046.

To test the consequences of a higher initial spin value of the BH, we ran a companion simulation to our fiducial simulation. The only difference between the two was that the companion simulation had an initial SMBH spin value of 0.8. As can be seen in the top panel of Fig. \ref{fig:spin}, the SMBH spin persistently decreases over the course of the simulation, until it converges with the fiducial simulation after $\sim 500$ Myr. While the spin is high, the jet changes direction very slowly in comparison to the fiducial simulation, as the high angular momentum of the rapidly spinning BH makes reorientation more difficult. Once the spin has dropped below 0.4, the jet direction changes more rapidly and the two simulations become statistically indistinguishable. The bubbles remain comparatively fat even in the absence of precession. This is due to the fact that our jets are very light and hot, and therefore over-pressurized in comparison to the background medium. While injected bimodally, the bubbles quickly expand outwards into the surrounding medium. We note that the absence of magnetic fields, whose wound-up helical structure along the jet is expected to keep it confined over kpc scales \citep[see][for a review]{Pudritz2012}, will have contributed to the fatness of the bubbles. We therefore postpone a comparison between bubble structures in a high spin and a low spin case to future, magnetised simulations.

\subsection{Dense gas structures}
\label{sec:clump_properties}

\subsubsection{Quantifying clump morphology}

As can be seen visually in Fig. \ref{fig:pretty_picture}, the dense gas in the cluster center can be found in clumps of a wide range of sizes and shapes. A clump is defined here to be a connected volume of space, for which all cells have a minimum density of 1 $\rm H\, \rm cm^{-3}$ and a maximum temperature of $10^6$ K. All properties are measured by summing over all cells contained within a given clump. Tracer particles are associated with a particular clump if they are contained within the clump volume at the point of measurement.

To quantify this parameter space, we measured the physical extent of individual clumps using the following methodology:
\begin{enumerate}
\item{Find the center of mass for each clump by summing over all cells contained within the clump, treating each cell as a point mass located at the cell center.}
\item{Calculate the clump's mass-weighted reduced inertia tensor using 
\begin{equation}
    I_{i,j}= \sum\limits_{n=1} \frac{m_n x_{n,i} x_{n,j}}{R^2_n}
    \label{eq:inertia}
\end{equation}
by summing over all cells $n$ contained within a clump, where $x_{n,i}$ is the ith coordinate of the nth cell within the clump, measured in the center of mass frame of the clump. $R_n$ is the nths cells distance from said center of mass, and $m_n$ is its gas mass.}
\item{Calculate the physical extent of the major axis $r_{\rm maj}$ by finding the largest distance between any two cell centers contained within the clump. To this value,  $\Delta x_{\rm min}$ is added to extrapolate from the cell centers to the cell edges contained within the clumps.}
\item{Find the axis vectors and axis length ratios using the eigenvalues and eigenvectors of the inertia tensor from Eq. \ref{eq:inertia}.}
\item{Calculate the median and minor axis length, $r_{\rm med}$ and $r_{\rm min}$ respectively, using the axis length ratios from the previous step, and the length of the major axis, $r_{\rm maj}$.}
\item{Calculate the volume of the ellipse defined by the three axes:
\begin{equation}
    V_{\rm ellipse}=\frac{4}{3}\pi r_{\rm maj} r_{\rm med} r_{\rm min}.
    \label{eq:vellipse}
\end{equation}}
\item{Calculate the volume filling fraction $f_{V}$, which is defined to be the ratio of the volume defined by the axis vectors, $V_{\rm ellipse}$ in Eq. \ref{eq:vellipse}, and the sum of the cell volumes contained within the clump:
\begin{equation}
    f_{V}=\frac{V_{\rm ellipse}}{\sum\limits_n V_n}
\end{equation}
where $V_{n}$ is the volume of the nth cell contained in the clump. For solid, round clumps well described by an ellipse, $f_{V}$ will have a value close to unity. For clumps with a complex morphology, such as bent filaments and three-dimensional networks of filaments and clumps, the volume fraction will be low as the axis vectors used to describe the ellipse mark the total physical extent of the clump along a given axis vector in 3D space, and said ellipse will therefore contain many cells outside the clump.}
\end{enumerate}

\begin{figure*}
    \centering
        \includegraphics[width=\textwidth]{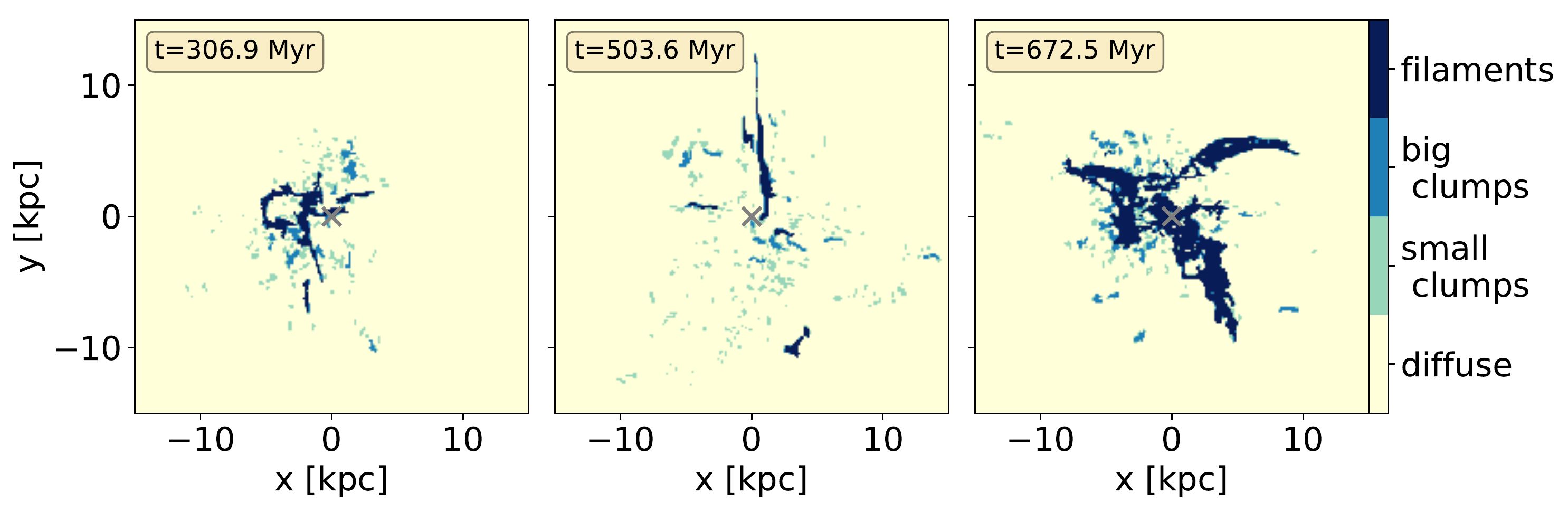}
    \caption{Example projections of the decomposition of structures into small clumps, big clumps and filamentary structures at three different points in time. Structures are considered to be distinct when not connected in 3D space. From left to right, the plots contain 5, 5 and 9 distinct filaments respectively, partially superimposed due to projection effects.}
    \label{fig:categorised}
\end{figure*}

For further analysis, we split the population of clumps into three categories depending on the length of their major axis relative to the mean major axis of the whole sample, $\bar{r}_{\rm maj} = 1.54$ kpc, and the samples standard deviation $\sigma_{\rm maj} = 1.42$ kpc:
\begin{enumerate}
    \item {\bf small clumps} have a major axis $r_{\rm maj}<\bar{r}_{\rm maj} =1.54$ kpc.
    \item {\bf big clumps} have a major axis length in the range $\bar{r}_{\rm maj} = 1.54  < r_{\rm maj} < \bar{r}_{\rm maj}+\sigma_{\rm maj}= 2.96$ kpc.
    \item {\bf filaments} have $r_{\rm maj} >  \bar{r}_{\rm maj}+\sigma_{\rm maj} = 2.96$ kpc.
\end{enumerate}
  Some example decompositions according to these criteria can be seen in Fig. \ref{fig:categorised}.
  
\subsubsection{Clump properties}

 \begin{figure*}
 % Made in http://localhost:8889/notebooks/2019_03_06_clump_morphologies.ipynb and in Paperplots
    \centering
    \includegraphics[width=\textwidth]{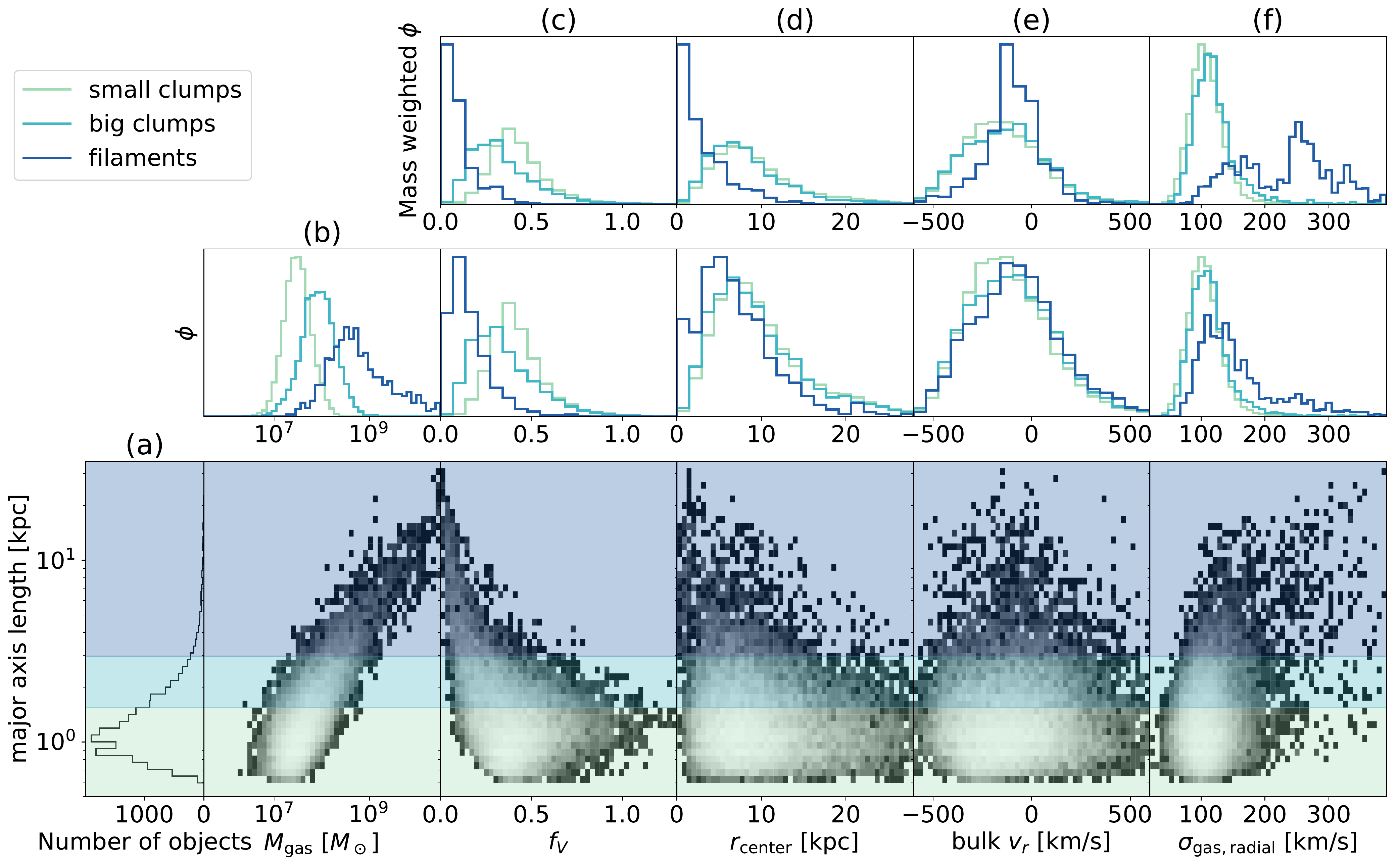}
    \caption{Clump properties for the whole sample (bottom row) and split into the three structure categories (top two rows).  From left to right: clump gas mass $M_{\rm gas}$, volume ratio $f_V$, distance between the clump center of mass and the cluster center $r_{\rm center}$, bulk velocity $v_r$ and gas velocity dispersion within the clump $\sigma_{\rm gas,radial}$. The probability distributions $\phi$ in the top row is mass weighted, while the one in the row below is unweighted.}
    \label{fig:clump_properties}
\end{figure*}

A variety of bulk clump properties versus axis length are shown in Fig. \ref{fig:clump_properties}, for the stacked sample of clumps of the whole simulation. As can be seen in column (a),  the distribution of major axis lengths ranges from the resolution limit of the simulation to very large, extended objects that have major axes of the order $10$ kpc or more. The stacked sample shown here, which contains all objects from all snapshots at all points in time of the simulation, contains 37897 small clumps (87.4~\%), 4283 big clumps (9.9~\%) and 1153 filaments (2.7~\%).

As expected, smaller clumps contain less gas mass (Fig. \ref{fig:clump_properties}, column (b)), with a minimum gas mass for the current resolution of $5 \times 10^5 \rm \ M_\odot$, and an average value of $1.8\times 10^7 \rm \ M_\odot$ for small clumps and $1.1 \times 10^8 \rm \ M_\odot$ for large clumps.  The population of filaments is much more massive, with an average gas mass of  $2.5 \times 10^9 \rm \ M_\odot$. Structures with a mass above $10^9 \rm \ M_\odot$ are all classified as filamentary. This lower mass limit for gas clumps is determined by the resolution. As we tested with a companion simulation, in which we reduced $\Delta x_{\rm min}$ to  30 pc, i.e. a factor 4 smaller than in the fiducial simulation. With this improved resolution, the gas structures fragment further into even smaller clumps, with a new minimum mass of $2.2 \times 10^3 \rm \ M_\odot$, and a new minimum axis length still approaching the resolution limit. This suggests that the shattering into smaller structures is by no means complete, and with even more resolution, the clumps would continue to break apart, as in the "cloudlet" model by \citet{McCourtetal18}. However, larger, filamentary structures continued to exist even in the higher resolution simulation.

In terms of shape, smaller clumps have higher values of $f_V$, so they are indeed much more compact (column (c), Fig \ref{fig:clump_properties}). Values of $f_V>1$ can occur for compact objects when the axis length for the median and minor axis are under-estimated in comparison to the true extent of the clump which happens mainly for clumps with less than 20 cells. However, the volume of the ellipse used to fit the clump never exceeds that of the sum of the cells contained in the clump by more than 40~\%. More extended objects have $f_V$ far below unity, which is an indicator of complex morphology. The most clumpy filament produced in this simulation still has $f_V<0.7$ so large clumpy structures do not form at any point of the simulation.

Small and big clumps have a similar radial distribution (column (d), Fig. \ref{fig:clump_properties}) and are preferentially found between $3-10$ kpc from the cluster center. Filaments, on the other hand, include both a subsample found at large radii, and a sample of particularly extended structures in the cluster center, an example of which can be seen in the right hand panel of Fig. \ref{fig:categorised}. This suggests that gas structures merge into larger objects as they reach the cluster center, consistent with a model in which small structures rain down onto a central massive gas structure. This structure can take the form of a massive gas disk, as for example seen in \cite{Li&Bryan14a} and briefly also in the simulation presented here (see Fig.~\ref{fig:gas_disc}), or in the form of an extended but not rotationally-supported object such as the one in the right hand panel of Fig. \ref{fig:categorised}, or the gas structures seen in the first, third and fourth snapshot of Fig. \ref{fig:pretty_picture}. 

In velocity space, all three populations are similarly distributed (column (e), Fig. \ref{fig:clump_properties}), with no discernible difference in the unweighted probability distribution of small and medium clumps, as well as filaments. The mass-weighted distribution in the top row shows that all three categories of structures are preferentially infalling (i.e. have $v_r<0$). The time-stacked sample of the simulation has an unweighted mean radial velocity of $75 \,\rm km\,s^{-1}$, with a full width half max of $198 \,\rm km\,s^{-1}$, where radial velocity is measured in 3D space with the SMBH at the origin. Negative values denote gas falling towards the SMBH. These values are comparable to observed bulk velocities of $ 100 \,\rm km\,s^{-1}$ but are at the upper end of observed velocity widths of $100 - 218 \ \rm km\,s^{-1}$ for molecular gas in Perseus \citep{Salome2008,Hitomi2016,Gendron2018}. By comparison, they fall easily within the range of observed velocity widths for warm ionised gas in massive clusters \citep{Hamer2016}. We note that, in contrast to the observational values, the full width-half max calculated here is calculated across the entire time-stacked sample, not just along the line of sight. While the mean and dispersion values show good agreement with observations, the sample of clumps presented here has an overall larger velocity range than found in cold-gas maps of nearby clusters, which report velocity values across the map in the range of $350 \,\rm km\,s^{-1}$ at most \citep{Olivares2019,Gendron2018}. 

The velocity dispersion $\sigma_{\rm gas,radial}$ is defined to be the velocity dispersion of the radial velocities of all resolution elements within an individual clump. It therefore quantifies the range of velocities found within an individual object. Clumpy structures, both small and big, have a low velocity dispersion (column (f), Fig \ref{fig:clump_properties}), i.e. a small range of radial velocities,  with an average value of just $90 \,\rm km\,s^{-1}$. The bulk of the filaments, despite major axis lengths of 10 kpc or more, have radial velocity dispersion of less than $200 \,\rm km\,s^{-1}$ but there is a small population of high-velocity dispersion objects with $\sigma_{\rm gas}>200 \,\rm km\,s^{-1}$, which is preferentially populated by filaments: They make up 28~\% of the high dispersion objects versus only 2.7~\%  of the total sample.

Dynamically, the clumps are therefore a surprisingly uniform population, despite more than 2 orders of magnitude in size difference, and more than 4 orders of magnitude in mass range. Gas properties across all three populations are also similar, with a temperature range of $10 - 10^6$ K (the latter being the cut-off temperature for the definition of a dense gas structure in this paper), and densities in the range of $1 - 10^5 \,\rm H\, cm^{-3}$. The bulk of the gas has a temperature around $10^4$ K and a density of $10-10^3 \,\rm H\, cm^{-3}$. This is not to say that all objects have the same properties at a given point in time, but that all types of objects can be found at all points in phase space at some point throughout the simulation. 

\begin{figure}
    \centering
    \includegraphics[width=\columnwidth]{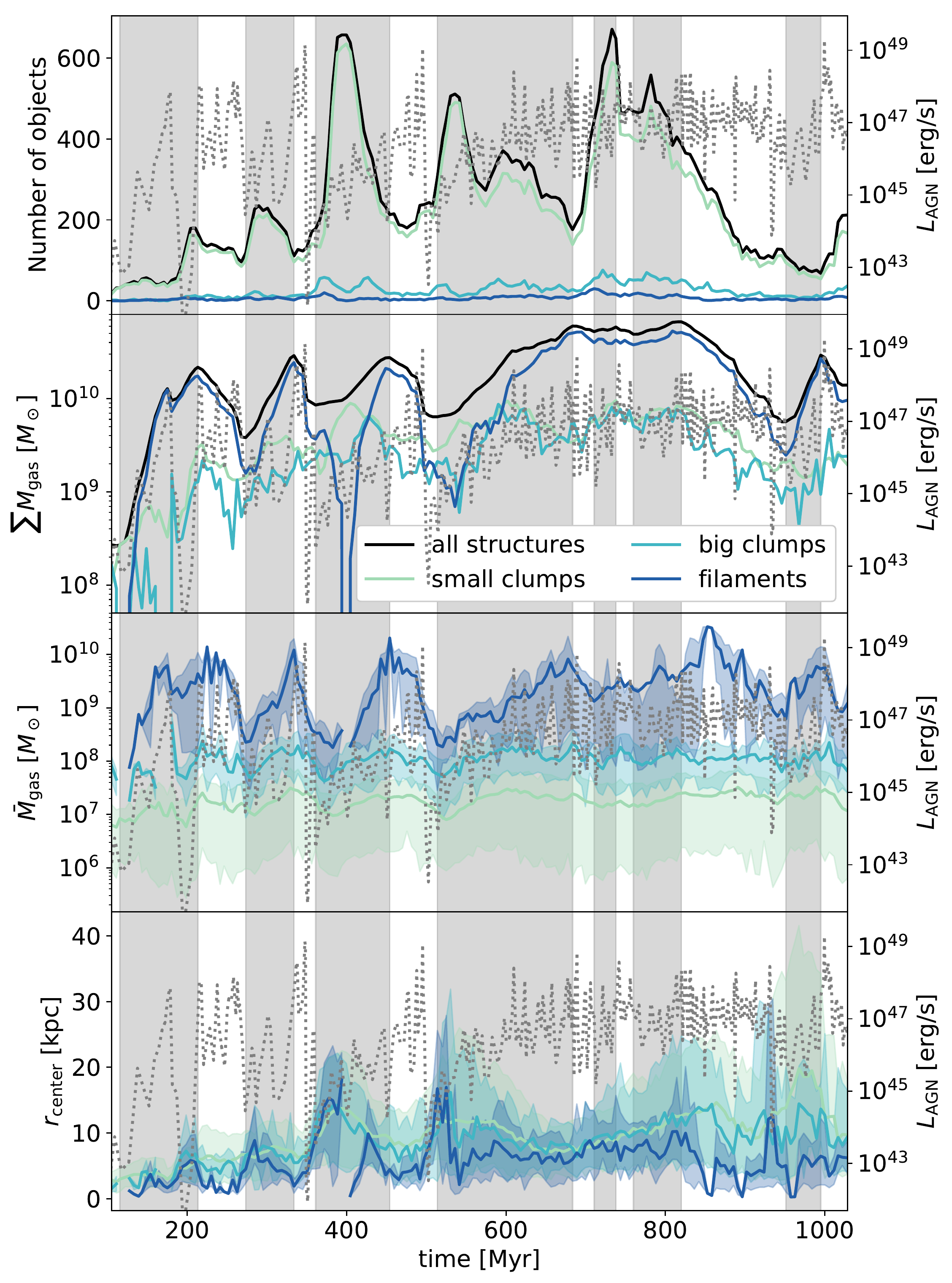}
    \caption{Time evolution of (from top to bottom) the number, total dense gas mass, mean gas mass and mean distance to the cluster center for the three structure categories. In the bottom two panels, solid lines show the mean and shaded regions the range from the 10th to the 90th percentile of the distribution. The dashed grey line in all plots shows the AGN luminosity for comparison.}
    \label{fig:decomposed_timeseries}
\end{figure}

The morphology and distribution of objects can vary strongly on a 5 Myr timescale, as can be seen in Fig. \ref{fig:decomposed_timeseries}. Overall, the number of structures at all points in time is dominated by small clumps, which are always the most abundant and make up 87.4~\% of the time-integrated sample. During some parts of the cooling-dominated phase, they also contain the bulk of the dense gas mass, such as around 400 Myr and at $500-550$ Myr. The rest of the time, the bulk of dense gas mass can be found in filaments, despite the fact that they only make up 2.7~\% of the overall sample by number. Big clumps contain dense gas mass on the order of that contained in the small clumps, but represent 9.9~\% of the total number of objects. 

From Fig. \ref{fig:decomposed_timeseries}, strong bursts of AGN feedback are followed by a strong increase in the number of small clumps, as well as an equally strong drop in both the total mass of gas contained in filaments (second panel) and the average mass of gas per filament (third panel). At the same time, the average radial distance between the cluster center and a clumps center of mass increases (bottom panel). While the bulk of clumps can usually be found within the central 20 kpc of the cluster, strong AGN outbursts produce clumps at much larger radii, up to 50 kpc from the location of the cluster center. This suggests a scenario where large objects are being shattered into smaller clumps during their interaction with strong AGN jets, and highlights the importance of the AGN jet not just for slowing down cooling onto the cluster center but also for the morphology and kinematics of the existing dense gas structures. The details of this interaction will be explored further in the next section.

\subsection{Uplifting}
\label{sec:uplifting}

Uplifting has been used to explain the unstructured velocity profiles observed in nearby clusters \citep{Pulido2018,Gendron2018}. When talking about uplifting dense gas in clusters, two different mechanisms need to be distinguished. On the one hand, there is the entrainment of existing dense gas by the AGN driven outflows, which turns previously infalling dense gas into outflowing dense gas, which will be discussed in this section. Alternatively, outflowing dense gas could form via condensation at large radii, when gas is uplifted from the cluster center by AGN jets, before being deposited at larger radii, where local entropy conditions then allow gas to condense \cite{Voit2017,Voit2018}.

\begin{figure*}
    \centering
    \subfloat[Visual time evolution of one episode of AGN feedback that starts around $t=350$ Myr. Only the dense gas is plotted. The colourmap shows the radial velocity of the gas, with negative values denoting infall, with the background color set to grey for clarity. The location of the BH is marked by a cross, and the contours show the extent of the AGN feedback bubbles produced by the feedback event that starts at $t=323$ Myr.]
    {\includegraphics[width=\textwidth]{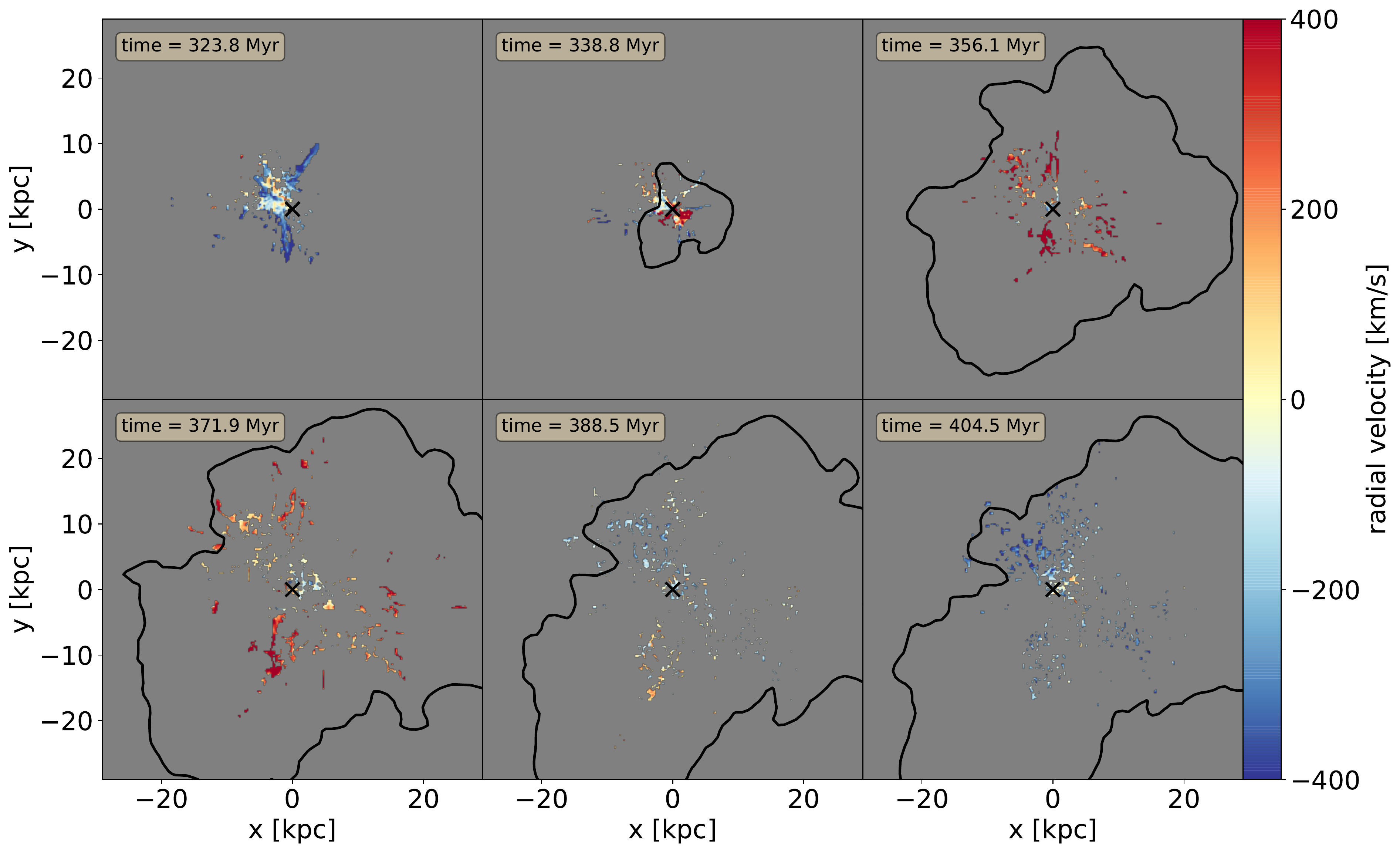}}
    \newline
    \subfloat[Time evolution of dense gas mass contained in the three categories over the same period of time. Vertical grey lines mark the outputs shown in the top panel.]
    {\includegraphics[width=\textwidth]{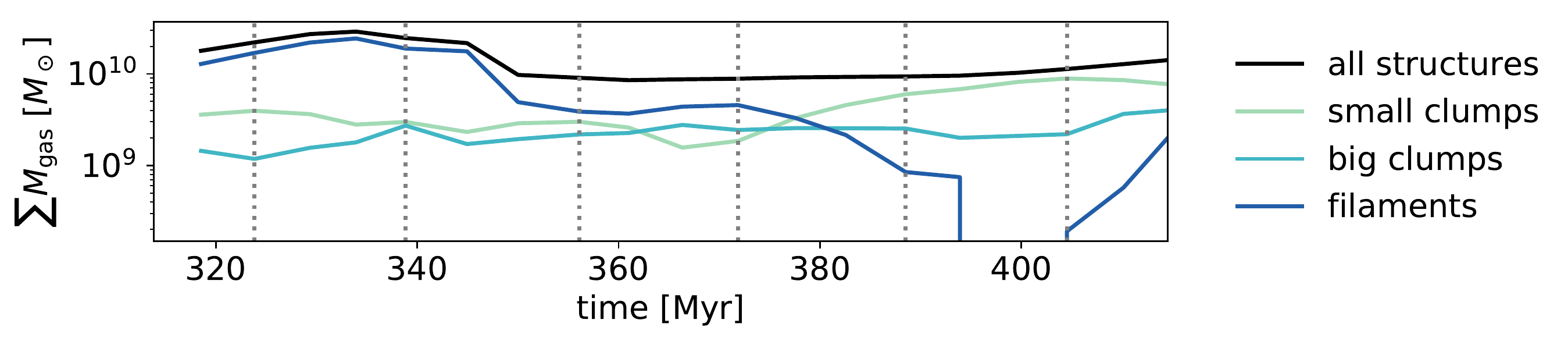}}
    
    \caption{Visual time evolution of one episode of AGN feedback that starts around $t=350$ Myr (top plot) and time evolution of dense gas mass contained in the three categories over the same period of time (bottom plot).}
    \label{fig:uplifting_image}
\end{figure*}

The impact of one interaction between the AGN jet and the dense gas in the cluster center, namely the outburst at $320 - 400$ Myr, is shown visually in Fig. \ref{fig:uplifting_image}: at $t=323$ Myr (top left), the dense gas is predominantly infalling and contained in  radially oriented filaments. At this point in time, the filaments contain ${M_{\rm gas,filaments} = 1.7 \times 10^{10} M_\odot}$, i.e. 76~\% of the total dense gas mass, with an average gas mass per filament of ${\bar{M}_{\rm gas,filaments}=1.47 \times 10^8 \rm \ M_\odot}$. As the AGN outburst commences, fed by this infalling dense gas ($t=338-356$ Myr, middle and top right panel), the filaments are broken into small and medium size clumps, and their velocity turns from infalling to outflowing. By ${371 \rm \ Myr}$ (bottom left), gas is predominatly outflowing, and the total mass budget of $8.8 \times 10^{10} \rm \ M_\odot$ is evenly split between small clumps, medium clumps and filaments. The filaments that continue to exist are much less massive, with an average mass of just $\bar{M}_{\rm gas,filaments} = 4.2 \times 10^7 \rm \ M_\odot$.

By $t=388$ Myr, the gas has reached its largest radial extent for this episode and is beginning to fall back onto the cluster center in the form of a shower of small, distinct clumps. From 371.9 Myr to 388.5 Myr, the total gas only increases by 5~\%, from $8.8 \times 10^{10} \rm \ M_\odot$ to $9.3 \times 10^{10} \rm \ M_\odot$, but the total number of objects triples as objects continue to break apart, from 244 at 371.9 Myr to 651 individual objects by 388.5 Myr. By this point, small clumps dominate the population, as they represent 94~\% of objects and contain 64~\% of the total gas mass, with a further ${27~\%}$ contained in big clumps. 

The timeseries of the number of different objects in the top panel of Fig. \ref{fig:decomposed_timeseries} shows that this behaviour is generic for the cluster presented here. Following a strong feedback outburst, the number of small objects spikes, while the total gas mass and the average mass per filament decrease strongly. At the same time, the average distance for objects of all categories increases as they are ejected from the cluster, with the outermost small clumps being found as far as 40 kpc or more from the cluster center. 

\begin{figure}
    \centering
    \includegraphics[width=\columnwidth]{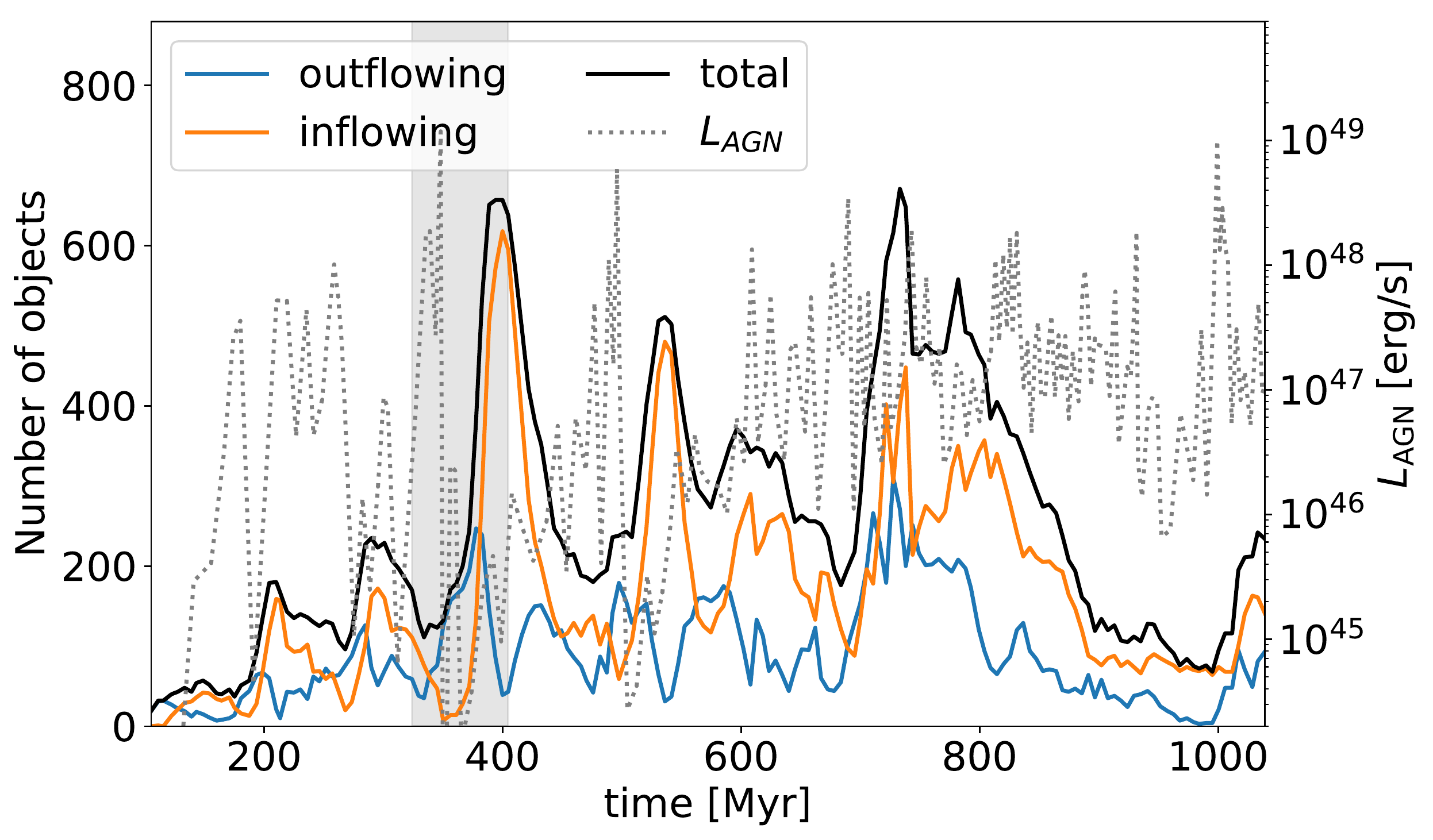}
    \caption{Time evolution of the total number of inflowing and outflowing clumps. The AGN luminosity is shown as a dotted line for comparison. The solid background highlights the event shown in Fig. \ref{fig:uplifting_image}.}
    \label{fig:clump_velocity_timeseries}
\end{figure}

Looking directly at the number of inflowing and outflowing objects, as shown in Fig. \ref{fig:clump_velocity_timeseries}, strong AGN feedback bursts are followed by a spike in the number of outflowing objects, as larger, filamentary structures are entrained and broken up by the hot winds of AGN feedback and lifted to larger radii. As gas is evacuated from the cluster center the AGN turns off. The entrained clumps then decelerate under gravity and fall back onto the cluster center. During this process, they shatter into even smaller components  so the number of individual objects continues to increase even after the AGN has become quiescent again. As the small clumps fall back onto the cluster center, they coalesce and trigger another strong outburst of AGN feedback, which repeats the cycle. The results presented in this paper are similar to work by \cite{Yang&Reynolds16}, who presented evidence for existing dense gas to be redistributed by the AGN jet. Contrary to their work, the dense gas in the simulations presented here is not indestructible. In our simulations, only 25~\% of the dense gas survives its interaction with the hot jet. It gets entrained by the AGN driven outflows and lifted to large radii. We note that, with a temperature cut of $10^6$ K, the gas discussed here is equivalent to the ionised dense gas seen in observation, not to the molecular gas. We expect that if we were able to adequately distinguish between ionised warm gas and molecular cold gas, the molecular gas would be much more difficult to uplift by the AGN jet.

This is surprising in the context of work by \citet{Klein1994}, who showed that for adiabatic cold structures in hot winds, the drag timescale $t_{\rm drag} \approx \chi r_{\rm clump} / v_{\rm wind}$ is always longer than the clump crushing timescale $t_{\rm cc} \approx \chi^{1/2}  r_{\rm clump} / v_{\rm wind}$, where $\chi$ is the density contrast between wind and cold clump, $r_{\rm clump}$ is the clump radius and $v_{\rm wind}$ is the relative velocity. It should therefore be impossible to accelerate cold clumps with a hot wind. However, recent work by \citet{Gronke2018} shows that radiative cooling can replenish the cold clump mass from the hot gas during uplifting and thereby dramatically increase the clump lifetime. Under these assumptions, clumps with radii larger than $r_{\rm clump} > v_{\rm wind} t_{\rm cool,mixing}/\chi$, where $t_{\rm cool,mixing}$ is the cooling time in the mixing layer surrounding the cold clumps, should survive the uplifting process, as cooling from the hot to the cold phase replenishes gas faster than cold gas from the clumps is being evaporated. For the simulation presented here, the maximum outflow velocities in the vicinity of clumps is of the order $10^4\,\rm km\,s^{-1}$, the cooling time in the mixing layer around clumps is of the order $0.1$ Myr and the density contrast $\chi \approx  10^{4}$. Therefore, clumps with a minimum value of $r_{\rm clump} \approx 1$ pc should survive their interaction with the hot wind, which is much smaller than the smallest cell size of 120 pc. While poorly resolved clumps most likely lack this mixing layer, and are therefore destroyed during the jet interaction, well-resolved cold clouds would be expected to survive their interaction with the hot outflows and become entrained without being destroyed, as shown in Fig. \ref{fig:uplifting_image}. These results are also in agreement with work by \cite{Armillotta2017}, who show that the bulk of cold gas in clouds with radii above 250 pc survives being accelerated by a hot wind for 200 Myr. It is  however likely that the 25~\% of dense gas that survives the interaction in our simulations is an overestimate, as work by \citet{Sparre2019} showed that more highly resolved clouds shatter more efficiently during their interaction with hot winds and therefore have shorter overall lifetimes than less resolved clouds.

\begin{figure*}
%Made using 2019_04_17_los_velocity_images.py
    \centering
    \includegraphics[width=\textwidth]{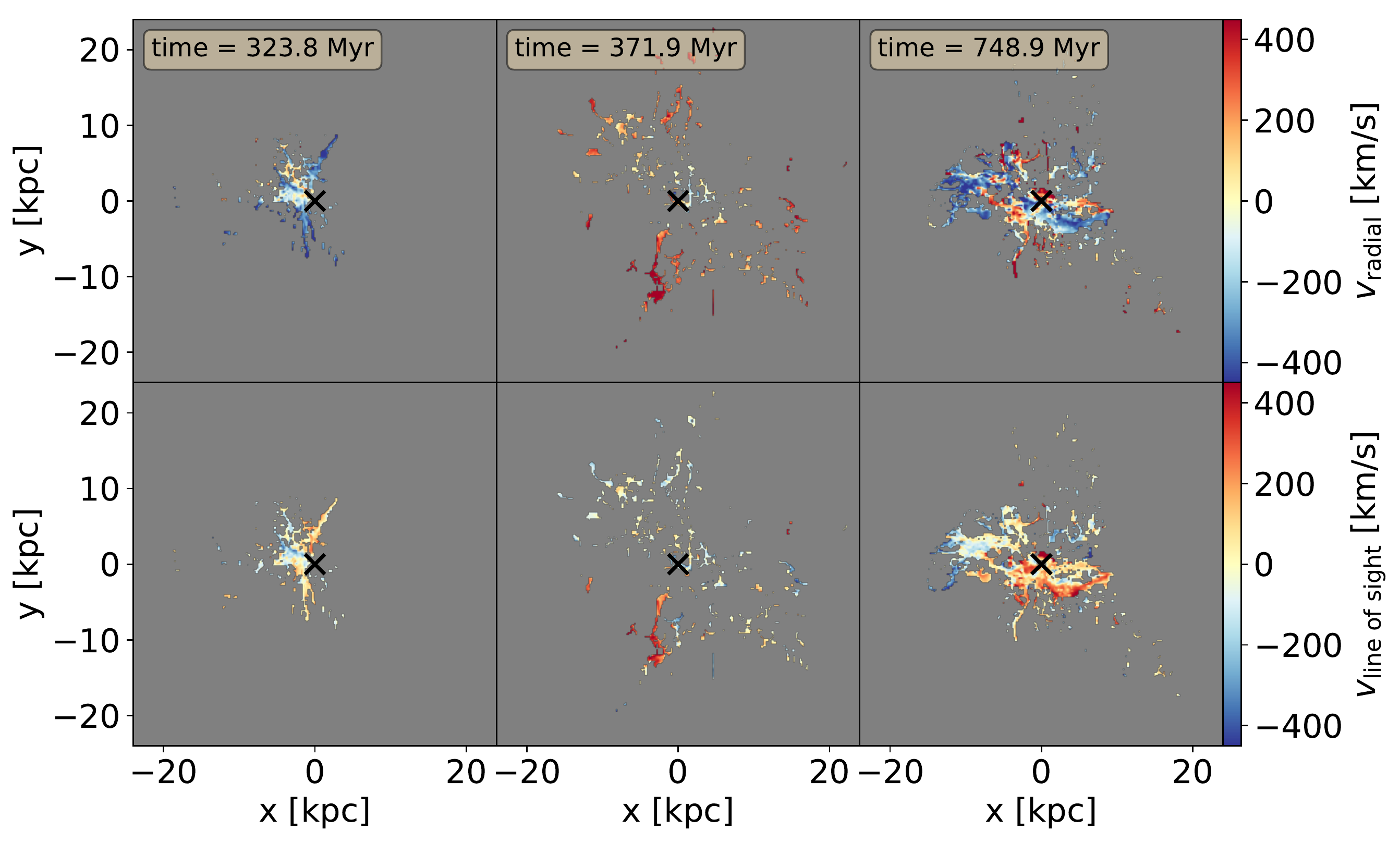}
    \caption{Density weighted velocity projections of the dense gas at three different points in time. The top row shows the radial velocity for each snapshot, the bottom row the corresponding line of sight velocity (here chosen to be the z-axis of the simulation box).}
    \label{fig:los_image}
\end{figure*}

\begin{figure*}
%Made using 2019_04_17_los_velocity_images.py
    \centering
    \includegraphics[width=\textwidth]{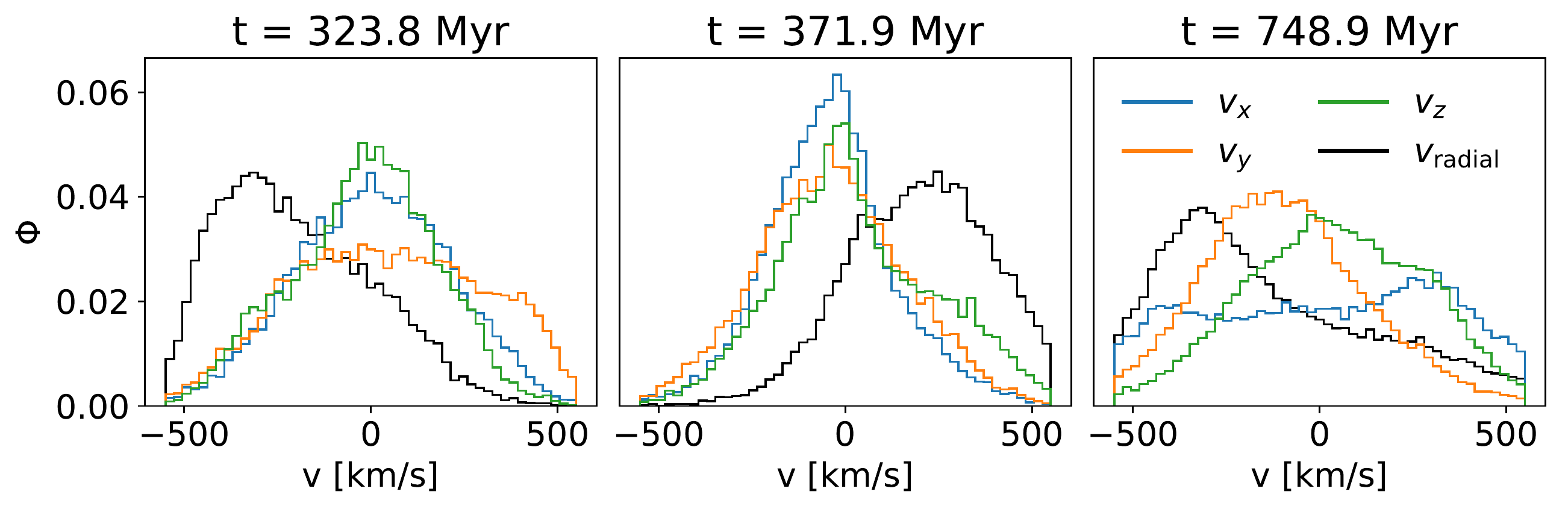}
    \caption{Distribution of resolution elements in radial velocity, and line of sight velocity along the x-axis, y-axis and z-axis of the simulation box respectively, for the three snapshots in time shown in Fig. \ref{fig:los_image}.}
    \label{fig:los_distribution}
\end{figure*}

In comparison to the observed velocity maps for H$\alpha$ emitting gas in Perseus by \cite{Gendron2018}, the velocity maps from our simulation (as shown in Fig. \ref{fig:uplifting_image}) are much more coherent, with clumps either predominantly infalling or outflowing in a given map. In this context we note that the maps in Fig. \ref{fig:uplifting_image} show an unusual period for our cluster, i.e. the only AGN outburst during which the  number of infalling clumps fall almost to zero (see Fig. \ref{fig:clump_velocity_timeseries}). This episode was chosen for analysis as it illustrates uplifting by AGN feedback particularly cleanly. At other points in time, dense gas can be observed to be inflowing and outflowing at the same time in our simulation, due to the directionality of the jet and the limited width of the jet cone.

It is also important to remember that the observed velocities are line-of-sight velocities, while Fig. \ref{fig:uplifting_image} shows radial velocities. As can be seen visually in Fig. \ref{fig:los_image}, which shows both radial velocities (top row) and line-of-sight velocities (bottom row) for an inflowing dominated (left column), an outflow dominated (middle panel) and a mixed (right column) point in time, the line-of-sight velocities appear less ordered than the radial velocities. The outflow or inflow dominated nature of the flow (left or middle panels respectively) cannot easily be recovered from line-of-sight velocity maps. This difficulty in distinguishing between flow patterns in the frame of the cluster, and line-of-sight flow patterns, is even more obvious in Fig. \ref{fig:los_distribution}, which shows the radial velocity probability distributions for the three snapshots in Fig. \ref{fig:los_image}, as well as that for the three line-of-sight velocities (here aligned with the $x$-axis, $y$-axis and $z$-axis of the box respectively). In all three cases, the line-of-sight velocities fail to recover the radial velocity pattern, even if the flow is clearly inflow or outflow dominated, and predict a more gaussian-like pattern with a mean velocity close to zero. The Gaussian distribution of line-of-sight velocities is expected for infalling or outflowing gas distributed roughly spherically around the cluster center. Therefore, the chaotic velocity patterns observed in nearby clusters are not necessarily evidence for the absence of coherent radial flow of the gas. 

% \TD{Who else saw results like these?
% \begin{enumerate}
%     \item Gaspari2017: Can we argue from their model that entrainment factors must be high, to prevent the BH from growing too efficiently when linking all the mass flow on different scales?
% \end{enumerate}}

\subsection{Condensation}
\label{sec:condensation}

As first proposed in \cite{McCourt2011}, and then shown in idealised simulations by \cite{Sharma2012}, dense gas can form out of the hot ICM via local thermal instability, even if the cluster is globally thermally stable. Condensation can happen when locally, $t_{\rm cool} / t_{\rm ff} $ falls below 1, and is suppressed for higher values. With sufficient uplifting of gas from the cluster center, condensation can occur for larger values of the radial $t_{\rm cool}/t_{\rm ff}$ profile, up to the range of $10-30$ \citep{Voit2017,Voit2018}, as also seen in observations \citep{Hogan2017,Pulido2018,Olivares2019}.

In the simulation presented here, we can use the tracer particles to estimate the condensation rate of dense gas. As each tracer particle has a unique identification number and traces $2 \times 10^4 \rm \ M_\odot $ of gas mass, the trajectories of tracer particles can be used to track gas flows throughout the simulation. The total mass of gas transferred from the hot, diffuse to dense phase between two simulation outputs can be estimated by counting the number of tracer particles that pass from the diffuse phase to the dense phase between two simulation outputs. The condensation rate $\dot{M}_{\rm condensed}$ is then found by dividing the newly condensed gas mass $M_{\rm condensed}$ by the time it took to assemble it.

\begin{figure*}
	\centering
	\begin{tabular}{cc}
		{\bf Condensation rate }&{\bf  Dense gas}\\
		\includegraphics[width=0.45\textwidth]{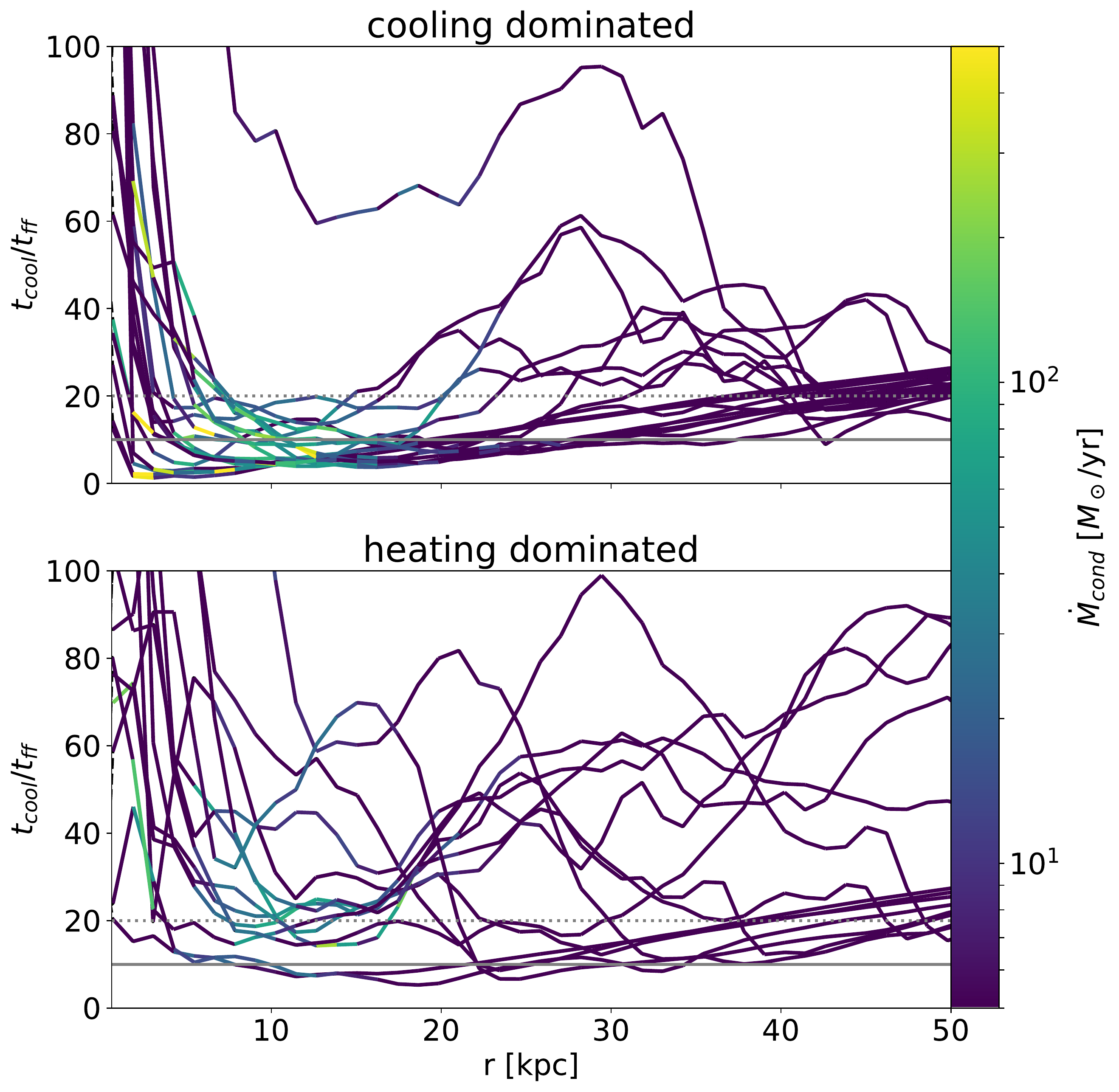} &
		\includegraphics[width=0.45\textwidth]{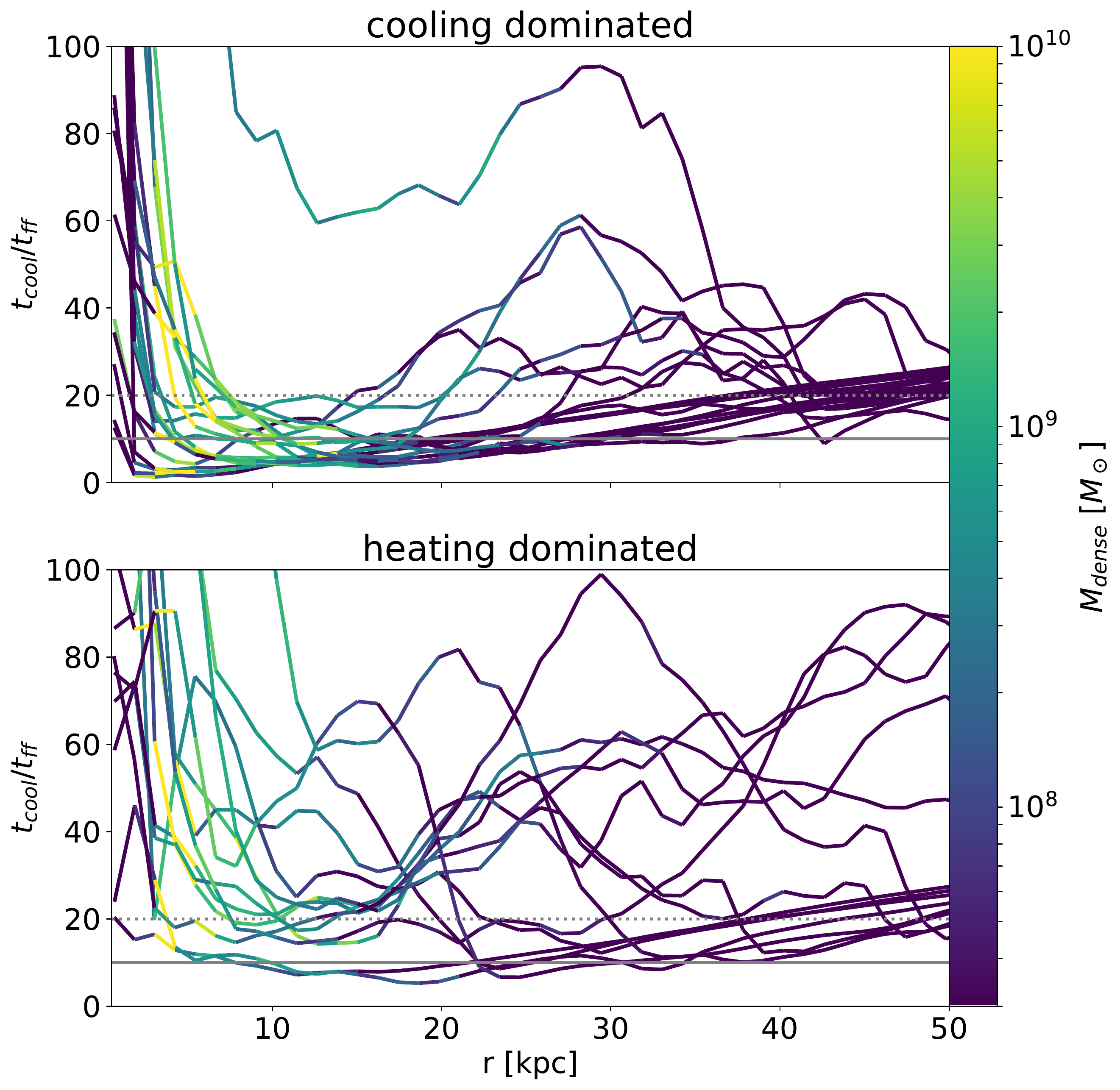}  \\
	\end{tabular}
	\caption{Cluster profiles of the cooling time ($t_{\rm cool}$) to free fall time ($t_{\rm ff}$) ratio at different snapshots of the simulation. The cluster profiles are sampled each 25 Myr across the full time evolution of the simulation. $t_{\rm cool}$ is calculated for each cell in the simulation, using its instantaneous density, temperature and cooling function as computed by RAMSES. $t_{\rm ff}$ is calculated using all mass (DM, gas, stars and the SMBH) contained within a given radius.  Profiles are colour-coded by condensation rate (left) or dense gas mass (right), based on the condensation rate and dense gas mass onto clumps at that radius. Snapshots during cooling dominated (top panel) and heating dominated intervals (bottom panel) are plotted separately for clarity.}
	\label{fig:tcooltff_profiles}
\end{figure*}

As can be seen in the left hand panels of Fig. \ref{fig:tcooltff_profiles}, our simulation confirms that condensation primarily occurs when $t_{\rm cool} / t_{\rm ff} < 20$. This is somewhat higher than prediction from idealised cooling simulations \citep{Sharma2012, McCourt2011}, most likely because the hot gas along the jet drives up the spherically averaged cooling time, but in line with observed values \citep{Hogan2017,Olivares2019}. Profiles of $t_{\rm cool}/t_{\rm ff}$ during the cooling dominated phases, which produce the bulk of the condensation, are generally ordered, with a clear minimum around 10 kpc. During heating dominated phases, by contrast, profiles show a much wider range of shapes as gas heated by the AGN rises to large radii in the form of hot bubbles, which significantly increase the cooling time both in the center and at larger radii. Some condensation continues during the heating dominated phases, and while the condensation remains confined to $<$ 20 kpc from the cluster centers, the values of $t_{\rm cool} / t_{\rm ff}$ can be as high as 50 even for actively cooling clusters. We postulate that this continued condensation is due to the multiphase structure of the ICM and the directionality of AGN feedback. Both $t_{\rm cool}$ and  $t_{\rm ff}$ are calculated for the hot ICM only, and it takes even strong AGN feedback bursts some time to reach large volume filling factors and shut off condensation completely.

\begin{figure}
    \centering
    \includegraphics[width=1\columnwidth]{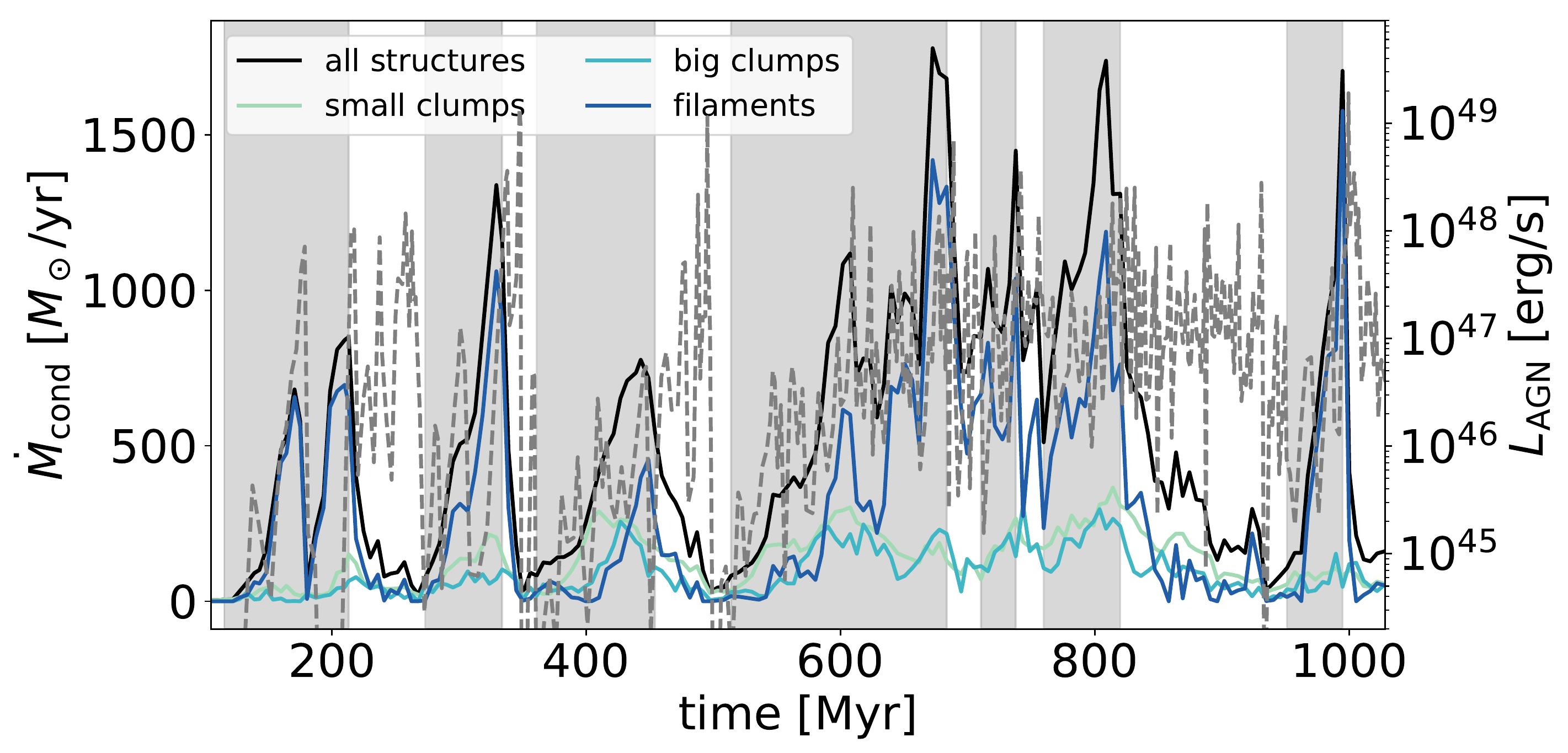}
    \caption{Time evolution for the gas condensation rate onto the dense structures in the simulation. See text for how the condensation rate is calculated.}
    \label{fig:decomposed_condensation}
\end{figure}

This hypothesis is confirmed by the condensation time-series in Fig. \ref{fig:decomposed_condensation}, which shows that condensation is highest towards the minimum of heating-dominated phases and falls to zero as the AGN feedback continues of impact the ICM. Fig. \ref{fig:decomposed_condensation} also shows that at the end of cooling-dominated phases, condensation occurs preferentially onto filamentary structures, but by the end of heating-dominated phases and the beginning of the next cooling-dominated phases, condensation occurs preferentially onto small and big clumps, in line with the uplifting - shattering - recondensation picture presented in Section \ref{sec:uplifting}.

As can be seen in Fig. \ref{fig:decomposed_condensation}, the total condensation rate of the cluster varies with time, ranging from a minimum of $3 \,\rm\ M_\odot\,yr^{-1}$ at the beginning of cooling dominated intervals to a maximum of up to $1.8 \times 10^3 \,\rm M_\odot\,yr^{-1}$ towards the end of cooling dominated phases. While the bulk of condensation takes place onto filaments, smaller and big clumps dominate when condensation rates are low. As discussed in the context of the clusters SFR in Section \ref{sec:cluster_evolution}, this condensation rate is high in comparison to the observed condensation rate for Perseus, which is in the range of $50 - 100\,\rm M_\odot$ \citep{Fabian2012}. In future work, we will explore if this over-cooling occurs because of the omission of non-thermal energies from cosmic rays in the work presented here, which are expected to be able to offset as much as 60~\% of the thermal cooling in a cluster environment \citep{Pfrommer2013,Jacob2017a,Jacob2017b, Ruszkowskietal17}.

While the areas of high condensation rate are confined to the minima of the $t_{\rm cool}/t_{\rm ff}$ profiles, dense gas can be found over a much wider range of radii (see righthand panels of Fig. \ref{fig:tcooltff_profiles}), and significant amounts of dense gas can also be observed during heating-dominated times. This is due to the fact that existing dense gas free-falls onto the cluster center from its formation location around 10 kpc, and is uplifted to larger radii due to its interactions with AGN feedback. The location at which dense gas is observed is therefore not a perfect proxy for where it is formed, as the kinematics in active clusters are complex and subject to hysteresis.

\begin{figure}
    \centering
    \includegraphics[width=\columnwidth]{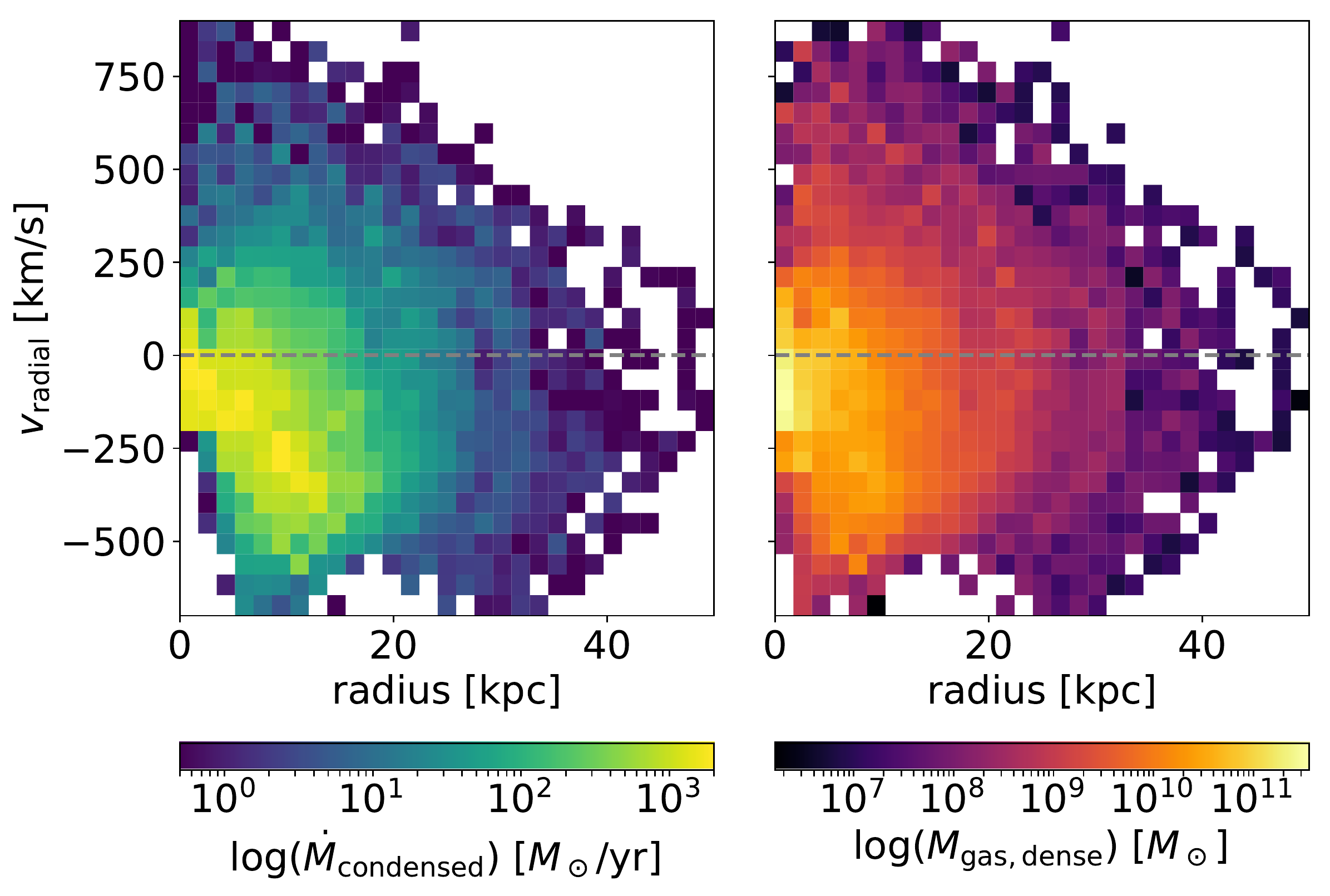}
    \caption{Phase plot of the total condensation rate (left) and total dense gas mass (right) over a range of radial positions and radial velocities of the clumps. Data shown here is stacked over all clumps at all snapshots of the simulation.}
    \label{fig:condensation_phaseplot}
\end{figure}

\begin{figure}
    \centering
    \includegraphics[width=\columnwidth]{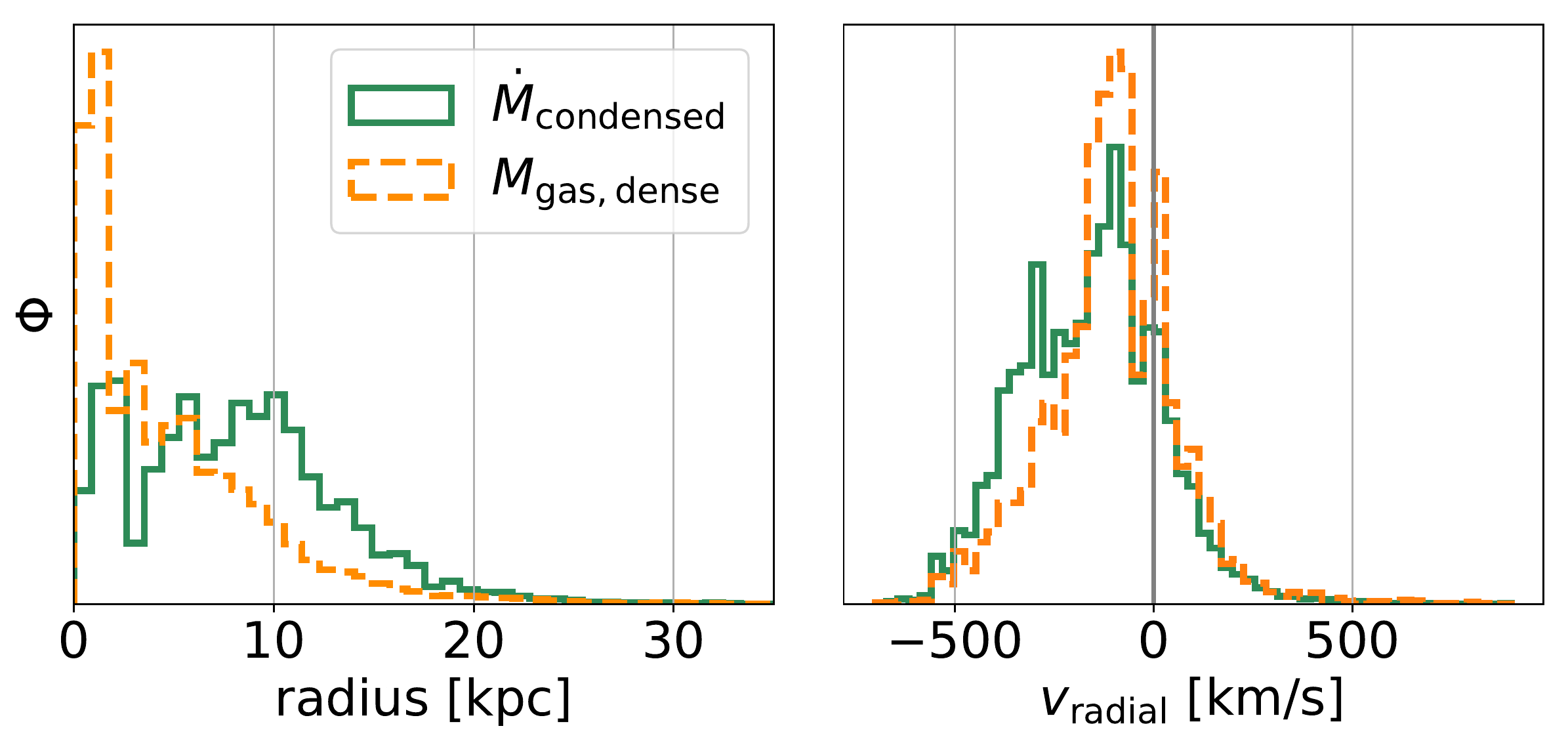}
    \caption{Probability distribution of clump radius (left)  and clump radial velocity (right) weighted by condensation rate and total dense mass respectively.}
    \label{fig:condensation_1D}
\end{figure}

This can be seen in more detail when comparing the radial and velocity distributions for stacked samples of newly condensed gas (left panel) and dense gas (right panel) in Fig. \ref{fig:condensation_phaseplot}. While some amount of condensation occurs over the full parameter space of radii and velocities occupied by dense clumps, the distribution in both radius and velocity is different for newly condensed gas and dense gas in general. As shown in both the mass distribution in Fig. \ref{fig:condensation_phaseplot}, and in the probability distributions in Fig. \ref{fig:condensation_1D}, dense gas is preferentially found at the cluster center, whereas condensation preferentially occurs at larger radii, with a peak of the distribution at 10 kpc. In velocity space, both existing dense gas and new condensation are preferentially infalling, but condensation has a broader distribution towards negative values, with a mean velocity at $-155.6 \rm \ km/s$ for condensation compared to $-104.3 \rm \ km/s$ for dense gas. Overall, only 75.9~\% of gas is infalling, while 82.1~\% of condensation occurs onto infalling clumps. This means that while the bulk of newly condensed gas is infalling, with an average condensation rate onto inflowing gas of $ 2.56 \,\rm M_\odot\,yr^{-1}$, there is also evidence for gas condensation onto outflowing clumps, which have an average condensation rate of $1.05\,\rm M_\odot\,yr^{-1}$.

We therefore conclude that condensation occurs preferentially onto infalling clumps within the radial range of $5-15$ kpc, but approximately a fifth of all condensation occurs onto outflowing gas.

\section{Discussion}
\label{sec:discussion}

In this paper, we have studied the formation, evolution and destruction of dense gas in the center of a Perseus-like cluster, under the influence of a spin-driven AGN jet, with a particular focus on the role of uplifting and condensation on the kinematics and morphology of the dense gas.

\subsection{Cooling and star formation}
\label{sec:sf_efficiency}
\begin{figure}
    \centering
    \includegraphics[width=\columnwidth]{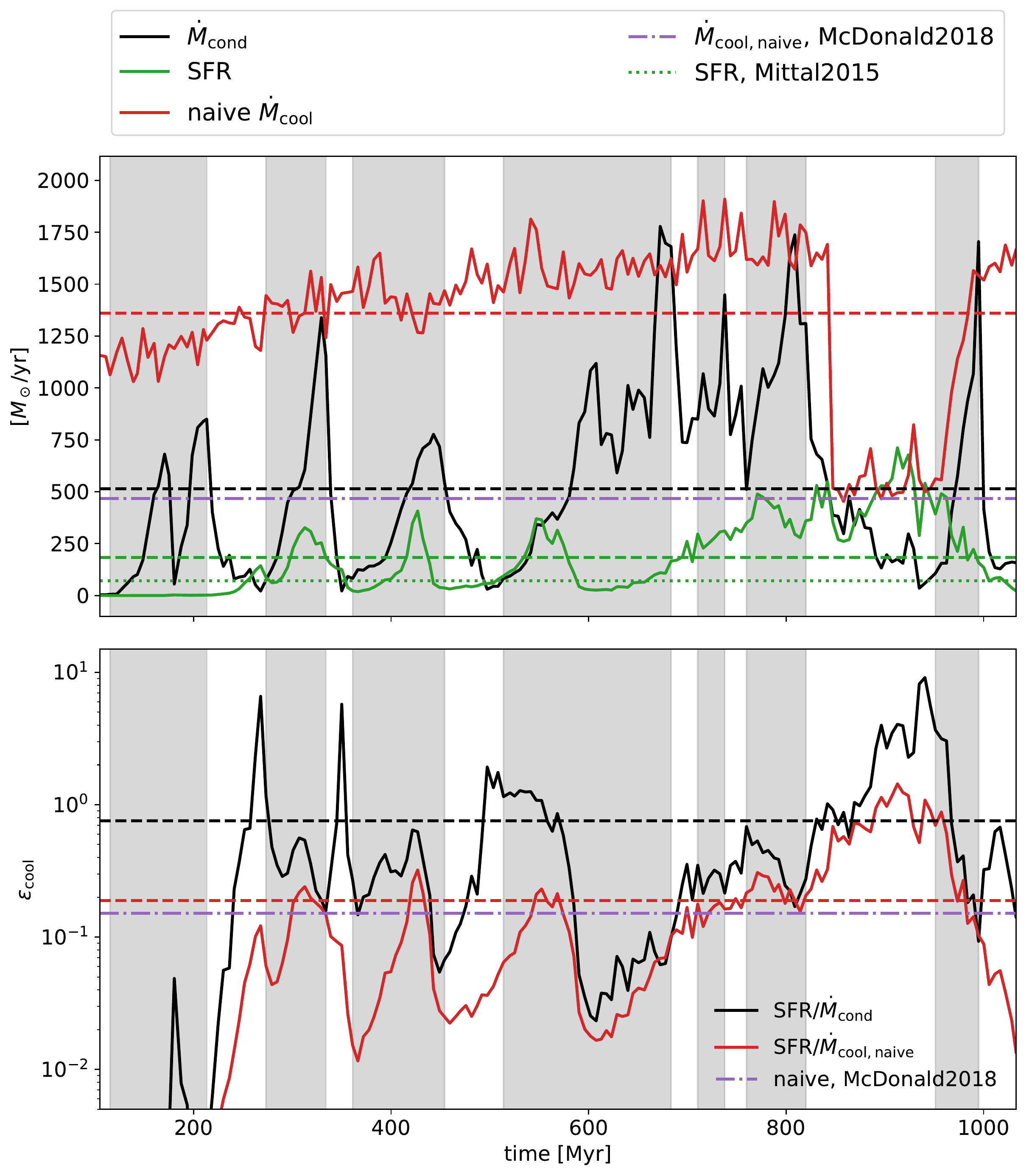}
    \caption{Cooling flow rates and star formation rates (top panel), as well as the resulting star formation efficiencies (bottom panel) using two different measures for the cooling rate, $\dot{M}_{\rm cond}$ and $\dot{M}_{\rm cool,naive}$. Solid lines show time evolution, while dashed lines show time averages. Dotted and Dashed-dotted lines show observational values for Perseus from \cite{Mittal2015}.}
    \label{fig:epsilon_sf}
\end{figure}
As reported in Section \ref{sec:cluster_evolution} and shown again in Fig. \ref{fig:epsilon_sf}, the star formation rate of the cluster (solid line) is much higher than observed values for Perseus, such as for example the 71 $M_\odot\,yr^{-1}$ measured by \citet{Mittal2015} (dotted line). As discussed in Section \ref{sec:cluster_evolution}, this could be due to an overly high cooling rate of the gas, or because too much of the resulting dense gas is turned into stars.

To compare the star formation efficiency of the cluster with observation, Fig. \ref{fig:epsilon_sf} shows both the total dense gas condensation rate $\dot{M}_{\rm cond}$ from Fig. \ref{fig:decomposed_condensation} and a naive cooling rate, defined as 
\begin{equation}
    \dot{M}_{\rm cool,naive}= \frac{M_{\rm gas}(r<r_{\rm cool})}{t_{\rm cool}(r_{\rm cool})}
\end{equation}
following \citet{McDonald2018}, where $M_{gas}(r<r_{\rm cool})$  is the total gas mass contained within the cooling radius $r_{\rm cool}$, which in turn is defined to be the radius at which the cooling time profile, $t_{\rm cool}(r)  = 3$ Gyr.

As can be seen in Fig. \ref{fig:epsilon_sf}, both the time series and the average value for $\dot{M}_{\rm cool,naive}$ are a factor 2-4 higher than the observed value, except during the disc-dominated stage between 850 - 950 Myr. By contrast, the dense gas condensation rate, $\dot{M}_{\rm cond}$ shows significant variablity but has a time-averaged value that is close to the naive observed cooling rate. It is also noticeably lower than the naive cooling rate, suggesting that reheating by the AGN keeps the majority of cooling gas from cooling efficiently and prevents it from condensating into dense gas.

Looking at the resulting star formation efficiencies ($\epsilon_{\rm cool}$, bottom panel), the average value of $\epsilon_{\rm cool} = {\rm SFR}/\dot{M}_{\rm cool,naive}=0.19 \pm 0.27$ is only slightly higher than the $\epsilon_{\rm cool} = 0.16$ reported by \citep{McDonald2018}, but the scatter on this value is large. The error on $\epsilon_{\rm cool}$ given here is equal to one standard deviation of the distribution. 

Looking at the efficiency of converting dense gas into stars, $\epsilon_{\rm cond} = {\rm SFR}/\dot{M}_{\rm cond}=0.76 \pm 1.37$ means that the  majority of dense gas is turned into stars. This shows that despite individual cold clumps loosing as much as 75\% of their mass during interactions with strong feedback episodes such as the one shown in Fig. \ref{fig:uplifting_image}, only about a quarter of the total dense gas is returned to the hot phase in this manner. Destroying dense gas once it has condensed is therefore not an efficient mechanism to regulate star formation in the cluster. Given the large variation in $\epsilon_{\rm cond}$, the instantaneous SFR is not a reliable tracer of $\dot{M}_{\rm cond}$, the cold gas formation rate of the cluster.

Fig. \ref{fig:epsilon_sf} also shows that in general $\dot{M}_{\rm cool,naive} >> \dot{M}_{\rm cond}$, so the vast majority of gas that cools out of the hot phase does not reach the dense phase, and is instead reheated by the AGN before condensing fully. As already known from the classic cooling flow problem  and confirmed here more quantitatively, $\dot{M}_{\rm cool,naive}$ is therefore not a good tracer of the overall cooling budget of the cluster as $\dot{M}_{\rm cond}/\dot{M}_{\rm cool,naive} = 0.38 \pm 0.27$ on average, with the error again denoting a standard deviation.

It is possible that we overestimate the star formation rate in dense gas, as we use a comparatively simple density-based star formation recipe of the form $\dot{\rho_*}=\epsilon_* \rho / t_{\rm ff}$, which does not take the effects of small-scale turbulence into account, and could therefore be too efficient for the context shown here \citep{Shi2011, Salome2016}.

\subsection{Filament lifetimes}
One notable result of our simulations is that extended gas structures form preferentially during comparatively AGN quiet times, and are readily destroyed in the interaction with AGN feedback. While this interaction between dense gas and AGN feedback is one of the requirements for effective self-regulation of cooling in the cluster, it also means that the lifetime of filaments is limited by the length of AGN duty cycles.

While as much as 25~\% of the dense gas mass survives the interaction with the hot, AGN driven outflows, larger structures are broken into smaller structures in the process. The result is a volume-filling distribution of small clumps, which are at first outflowing and then fall back onto the cluster center. Such a clumpy morphology of the dense gas is not supported by observations, which show more extended, filamentary structures \citep{Conselice2001,Fabian2005}. Two possible explanations come to mind.

One possibility is that the dense filaments are too readily destroyed in our simulations. If physical processes not modelled here, such as notably magnetic fields, could support the filaments against fragmentation, they might survive their interaction with the AGN jet and retain their extended morphology for longer. This hypothesis is supported by work on the survival rate of isolated clumps accelerated by hot, magnetised winds \citep{Shin08,McCourt2015,Xu2019}, which show that magnetised winds draw spherical clouds out into extended, filamentary structures instead of evaporating them or breaking them into smaller clumps. From this point of view, we over-estimate the fragmentation rate of dense filaments into the hot ICM.

The other possibility is that we underestimate the ability of AGN feedback to destroy dense clumps, for example by under-resolving the mixing layers at the outer clump surface \citep{Gronke2018}, or simply due to lack of resolution to follow the fragmentation process to smaller scales. This hypothesis is supported by our high-resolution companion simulation, which showed that the fraction of dense gas that survives this particular uplifting event falls from 25~\% at a resolution of $\Delta x_{\rm min}=120 $ pc to 19~\% at a resolution of $\Delta x_{\rm min}=30$~pc. The fact that the minimum clump size remains at the resolution limit shows that this process is by no means converged, and higher resolution would likely lead to even smaller clumps and even lower dense gas survival rates. This question has been investigated further by \citet{McCourtetal18}, who report that for individual clouds accelerated by a hot wind, even a sub-pc scale resolution is insufficient for fragmentation to converge. Based on work by \citet{Armillotta2017}, the survival rates for small gas clumps in hot winds is very low, which suggests that we would expect the gas currently contained in our small, compact gas clumps to break into an even large number of even smaller clumps until it evaporates entirely and mixes back into the ICM. From this point of view, we are under-estimating the fragmentation rate of small clumps, as well as under-estimating the ability of the AGN to evaporate dense gas.

\subsection{The width of filaments}
Throughout this paper, we have shown that extended dense gas structures readily form in the cluster center. While our filamentary dense gas structures show maximal extents of 1-10 kpc, in agreement with observations \citep{Conselice2001}, many of our structures appear much wider than the observed 70 pc.

Resolution will play a role in determining the width of the filaments, particularly for very thin filaments which currently have a width close to the resolution limit, such as the long, thing structures seen in the left two panels of Fig. \ref{fig:categorised}. A comparison simulation with higher resolution of $\Delta x_{\rm min} = 30$ pc, run for only a span of 15 Myr, produced thinner filaments than the fiducial simulation at 120 pc. However, many filaments seen in the fiducial simulation, such as for example the extended structures in the right hand panel of Fig. \ref{fig:categorised}, are well resolved at the current resolution and therefore not influenced by improvements in resolution. 

One process not modelled here, which is thought to play an important role in the morphology of filaments, is anisotropic thermal conduction along magnetic field lines. In the presence of anisotropic thermal conduction, in combination with magnetic fields, the characteristic thermal collapse length scale (the field length) becomes much larger along field lines than perpendicular to it \citep{Field1965}, as thermal energy is preferentially redistributed along field lines. Collapse therefore  preferentially occurs perpendicular to magnetic field lines, smearing spherical collapse out along magnetic field lines. Isolated simulations have shown that in the presence of magnetic fields, local thermal instabilities do indeed produce more extended filamentary gas structures \citep{McCourt2011,Ji2018,Xu2019} compared to more clumpy dense gas for the purely hydrodynamical runs. While this process could help smear dense, round clumps into long, extended filaments, it is unlikely to make the existing filaments thinner. Understanding why the filaments reported here take their particular shapes, and how their morphology might change in the presence of magnetic fields and cosmic rays, will be the subject of future work. 

Another limitation of our work is that with many structures shown here at the resolution limit of the simulation, it will be impossible to resolve the detailed internal structure observed for filaments, which consist of dense molecular clumps surrounded by an H-$\alpha$ envelope \citep{Salome2006,Salome2011}.
%Add cooling finction
With a more complex internal structure and gas dynamics, we would expect the energy balances of filaments to change, with as of yet poorly understood consequences for their morphology. 

\section{Conclusions}
\label{sec:conclusions}

In this paper, we have investigated the formation and evolution of dense gas in the center of a Perseus-like cluster under the influence of a spin driven AGN jet, using hydrodynamical simulations.

We showed that:  
\begin{enumerate}
    \item Under the influence of the AGN jet, the cluster undergoes repeated cycles of cooling dominated phases, when dense gas builds up in the cluster center, and heating dominated phases, when the total amount of dense gas decreases. Cycle lengths are on the order of 100 Myr, but show significant variation. (Section \ref{sec:cluster_evolution})
    \item For low black hole spin values, the chaotic cold accretion onto the cluster center is able to continuously reorient the spin axis, with characteristic reorientation timescales of the order of 10 Myr, allowing the jet to sweep out the full parameter space in both polar and azimuthal angle. (Section \ref{sec:jet}) 
    \item The morphology of dense gas is highly variable throughout the simulation, with between 20 and 620 individual dense structures present at a given point in time.  (Section \ref{sec:clump_properties})
    \item Major axis lengths of individual clumps range from the resolution limit of the simulation up to more than 30 kpc. Larger clumps have more complex, filamentary morphologies than smaller objects, which tend to be rounder and compacter. (Section \ref{sec:clump_properties})
    \item We find evidence for uplifting of existing dense gas by the AGN, with dense gas defined to have a maximum temperature of $10^6 \rm \ K$. During a strong feedback episode, larger, infalling structures fragment into smaller clumps under the influence of the hot outflows driven by the AGN. In the process, they lose up to 75~\% of their gas mass and become entrained and ejected from the cluster center. (Section \ref{sec:uplifting})
    \item Despite these high mass loss rates for individual clumps, 75.7 \% of the total dense gas is turned into stars during the course of the simulations. Despite individual clumps loosing up to 75 \% of their mass during interactions with AGN jets, destruction of dense gas via AGN feedback is therefore not an efficient channel to regulate star formation in clusters. (Section \ref{sec:sf_efficiency})
    \item A second round of fragmentation into even smaller clumps occurs at the top of the ballistic orbit, before surviving dense clumps fall back onto the cluster center where they re-coalesce into larger objects. (Section \ref{sec:uplifting})
    \item Condensation takes place preferentially when $t_{\rm cool}/t_{\rm ff}<20$, which occurs primarily during cooling dominated phases of the cluster, and in the radial range of $5-15 \rm \ kpc$. Heating dominated phases see more disturbed profiles of $t_{\rm cool}/t_{\rm ff}$ without a clear minimum as the ICM is heating by the AGN feedback. (Section \ref{sec:condensation})
    \item Dense gas continues to be observable even during heating-dominated phases, and is preferentially found at smaller radii than condensation, i.e. at $r<5 \rm \ kpc$, but can be found as far out as $30 \rm \ kpc$ due to uplifting. The presence of dense gas is therefore not a reliable tracer for condensation. (Section \ref{sec:condensation})
    \item While 82.1~\% of condensation of gas from the hot ICM onto dense clumps occurs on infall, there is also evidence for continued condensation for outflowing gas, with outflowing dense clumps having an average condensation rate of $1.06 \rm\ M_\odot\,yr^{-1}$, compared to $2.56 \rm\ M_\odot\,yr^{-1}$ for infalling clumps. (Section \ref{sec:condensation})
    \item Both direct uplifting of dense gas and condensation of gas from the hot, diffuse to the dense phase in outflowing gas has been invoked to explain the unstructured velocity maps observed in nearby clusters. While we find evidence for both mechanisms, and confirm a general lack of rotation in the dense gas, we also caution that the observed line of sight velocities fail to show coherent radial flow patterns even when they are present in the dense gas (Section \ref{sec:uplifting}).
    \item Neither the naive cooling rate $\dot{M}_{\rm cool,naive}$ nor the SFR are  reliable observational tracers of the cold gas formation rate $\dot{M}_{\rm cond}$. (Section \ref{sec:sf_efficiency})
\end{enumerate}

\begin{acknowledgements}
 The authors thank Maxime Trebitsch and Marta Volonteri for useful discussion, and the referee, Mark Voit, for his excellent comments and useful input. This work was supported by the ANR grant LYRICS (ANR-16-CE31-001 1) and was granted access to the HPC resources of CINES under the allocation A0040406955 made by GENCI. This work has made use of the Horizon Cluster hosted by Institut d'Astrophysique de Paris. We thank St\'ephane Rouberol for smoothly running this cluster for us. Visualisations in this paper were produced using the \sc{yt project} \citep{Turk2011}

\end{acknowledgements}

%%%%%%%%%%%%%%%%%%%%%%%%%%%%%%%%%%%%%%%%%%%%%%%%%%



\bibliographystyle{aa}
\bibliography{references} % if your bibtex file is called example.bib

\begin{thebibliography}{102}
\expandafter\ifx\csname natexlab\endcsname\relax\def\natexlab#1{#1}\fi

\bibitem[{{Agertz} {et~al.}(2013){Agertz}, {Kravtsov}, {Leitner}, \&
  {Gnedin}}]{Agertz2013}
{Agertz}, O., {Kravtsov}, A.~V., {Leitner}, S.~N., \& {Gnedin}, N.~Y. 2013,
  \apj, 770, 25

\bibitem[{{Armillotta} {et~al.}(2017){Armillotta}, {Fraternali}, {Werk},
  {Prochaska}, \& {Marinacci}}]{Armillotta2017}
{Armillotta}, L., {Fraternali}, F., {Werk}, J.~K., {Prochaska}, J.~X., \&
  {Marinacci}, F. 2017, \mnras, 470, 114

\bibitem[{{Beckmann} {et~al.}(2019){Beckmann}, {Devriendt}, \&
  {Slyz}}]{Beckmann2018b}
{Beckmann}, R.~S., {Devriendt}, J., \& {Slyz}, A. 2019, \mnras, 483, 3488

\bibitem[{{Bridges} \& {Irwin}(1998)}]{Bridges1998}
{Bridges}, T.~J. \& {Irwin}, J.~A. 1998, \mnras, 300, 967

\bibitem[{{Brown} {et~al.}(2011){Brown}, {Emerick}, {Rudnick}, \&
  {Brunetti}}]{Brown2011}
{Brown}, S., {Emerick}, A., {Rudnick}, L., \& {Brunetti}, G. 2011, \apjl, 740,
  L28

\bibitem[{{Cadiou} {et~al.}(2019){Cadiou}, {Dubois}, \&
  {Pichon}}]{Cadiouetal19}
{Cadiou}, C., {Dubois}, Y., \& {Pichon}, C. 2019, \aap, 621, A96

\bibitem[{{Canning} {et~al.}(2010){Canning}, {Fabian}, {Johnstone}, {Sanders},
  {Conselice}, {Crawford}, {Gallagher}, \& {Zweibel}}]{Canning2010}
{Canning}, R.~E.~A., {Fabian}, A.~C., {Johnstone}, R.~M., {et~al.} 2010,
  \mnras, 405, 115

\bibitem[{{Canning} {et~al.}(2014){Canning}, {Ryon}, {Gallagher}, {Kotulla},
  {O'Connell}, {Fabian}, {Johnstone}, {Conselice}, {Hicks}, {Rosario}, \&
  {Wyse}}]{Canning2014}
{Canning}, R.~E.~A., {Ryon}, J.~E., {Gallagher}, J.~S., {et~al.} 2014, \mnras,
  444, 336

\bibitem[{{Cattaneo} \& {Teyssier}(2007)}]{Cattaneo07}
{Cattaneo}, A. \& {Teyssier}, R. 2007, \mnras, 376, 1547

\bibitem[{{Choudhury} \& {Sharma}(2016)}]{Choudhury2016}
{Choudhury}, P.~P. \& {Sharma}, P. 2016, \mnras, 457, 2554

\bibitem[{{Cielo} {et~al.}(2018){Cielo}, {Babul}, {Antonuccio-Delogu}, {Silk},
  \& {Volonteri}}]{Cieloetal18}
{Cielo}, S., {Babul}, A., {Antonuccio-Delogu}, V., {Silk}, J., \& {Volonteri},
  M. 2018, \aap, 617, A58

\bibitem[{{Conselice} {et~al.}(2001){Conselice}, {Gallagher}, \&
  {Wyse}}]{Conselice2001}
{Conselice}, C.~J., {Gallagher}, John~S., I., \& {Wyse}, R. F.~G. 2001, \aj,
  122, 2281

\bibitem[{{Crawford} \& {Fabian}(1992)}]{Crawford1992}
{Crawford}, C.~S. \& {Fabian}, A.~C. 1992, \mnras, 259, 265

\bibitem[{{Dubois} {et~al.}(2010){Dubois}, {Devriendt}, {Slyz}, \&
  {Teyssier}}]{Duboisetal10}
{Dubois}, Y., {Devriendt}, J., {Slyz}, A., \& {Teyssier}, R. 2010, \mnras, 409,
  985

\bibitem[{{Dubois} {et~al.}(2011){Dubois}, {Devriendt}, {Teyssier}, \&
  {Slyz}}]{Dubois2011}
{Dubois}, Y., {Devriendt}, J., {Teyssier}, R., \& {Slyz}, A. 2011, \mnras, 417,
  1853

\bibitem[{{Dubois} {et~al.}(2014){Dubois}, {Volonteri}, {Silk}, {Devriendt}, \&
  {Slyz}}]{Duboisetal14bhspin}
{Dubois}, Y., {Volonteri}, M., {Silk}, J., {Devriendt}, J., \& {Slyz}, A. 2014,
  \mnras, 440, 2333

\bibitem[{{Ebeling} {et~al.}(1996){Ebeling}, {Voges}, {Bohringer}, {Edge},
  {Huchra}, \& {Briel}}]{Ebeling1996}
{Ebeling}, H., {Voges}, W., {Bohringer}, H., {et~al.} 1996, \mnras, 281, 799

\bibitem[{{Fabian}(1994)}]{Fabian1994}
{Fabian}, A.~C. 1994, \araa, 32, 277

\bibitem[{{Fabian}(2012)}]{Fabian2012}
{Fabian}, A.~C. 2012, \araa, 50, 455

\bibitem[{{Fabian} {et~al.}(2008){Fabian}, {Johnstone}, {Sanders}, {Conselice},
  {Crawford}, {Gallagher}, \& {Zweibel}}]{Fabian2008}
{Fabian}, A.~C., {Johnstone}, R.~M., {Sanders}, J.~S., {et~al.} 2008, \nat,
  454, 968

\bibitem[{{Fabian} {et~al.}(2003){Fabian}, {Sanders}, {Crawford}, {Conselice},
  {Gallagher}, \& {Wyse}}]{Fabian2003}
{Fabian}, A.~C., {Sanders}, J.~S., {Crawford}, C.~S., {et~al.} 2003, \mnras,
  344, L48

\bibitem[{{Fabian} {et~al.}(2006){Fabian}, {Sanders}, {Taylor}, {Allen},
  {Crawford}, {Johnstone}, \& {Iwasawa}}]{Fabian2005}
{Fabian}, A.~C., {Sanders}, J.~S., {Taylor}, G.~B., {et~al.} 2006, \mnras, 366,
  417

\bibitem[{{Fabian} {et~al.}(2016){Fabian}, {Walker}, {Russell}, {Pinto},
  {Canning}, {Salome}, {Sand ers}, {Taylor}, {Zweibel}, \&
  {Conselice}}]{Fabian2016}
{Fabian}, A.~C., {Walker}, S.~A., {Russell}, H.~R., {et~al.} 2016, \mnras, 461,
  922

\bibitem[{{Field}(1965)}]{Field1965}
{Field}, G.~B. 1965, \apj, 142, 531

\bibitem[{{Gaspari} {et~al.}(2011){Gaspari}, {Melioli}, {Brighenti}, \&
  {D'Ercole}}]{Gasparietal11}
{Gaspari}, M., {Melioli}, C., {Brighenti}, F., \& {D'Ercole}, A. 2011, \mnras,
  411, 349

\bibitem[{{Gaspari} {et~al.}(2013){Gaspari}, {Ruszkowski}, \&
  {Oh}}]{Gasparietal13}
{Gaspari}, M., {Ruszkowski}, M., \& {Oh}, S.~P. 2013, \mnras, 432, 3401

\bibitem[{{Gaspari} {et~al.}(2012){Gaspari}, {Ruszkowski}, \&
  {Sharma}}]{Gasparietal12}
{Gaspari}, M., {Ruszkowski}, M., \& {Sharma}, P. 2012, \apj, 746, 94

\bibitem[{{Gendron-Marsolais} {et~al.}(2018){Gendron-Marsolais},
  {Hlavacek-Larrondo}, {Martin}, {Drissen}, {McDonald}, {Fabian}, {Edge},
  {Hamer}, {McNamara}, \& {Morrison}}]{Gendron2018}
{Gendron-Marsolais}, M., {Hlavacek-Larrondo}, J., {Martin}, T.~B., {et~al.}
  2018, \mnras, 479, L28

\bibitem[{{Gnedin} \& {Hollon}(2012)}]{Gnedinetal12}
{Gnedin}, N.~Y. \& {Hollon}, N. 2012, \apjs, 202, 13

\bibitem[{{Gronke} \& {Oh}(2018)}]{Gronke2018}
{Gronke}, M. \& {Oh}, S.~P. 2018, \mnras, 480, L111

\bibitem[{{Hamer} {et~al.}(2014){Hamer}, {Edge}, {Swinbank}, {Oonk}, {Mittal},
  {McNamara}, {Russell}, {Bremer}, {Combes}, \& {Fabian}}]{Hamer2014}
{Hamer}, S.~L., {Edge}, A.~C., {Swinbank}, A.~M., {et~al.} 2014, \mnras, 437,
  862

\bibitem[{{Hamer} {et~al.}(2016){Hamer}, {Edge}, {Swinbank}, {Wilman},
  {Combes}, {Salom{\'e}}, {Fabian}, {Crawford}, {Russell}, \&
  {Hlavacek-Larrondo}}]{Hamer2016}
{Hamer}, S.~L., {Edge}, A.~C., {Swinbank}, A.~M., {et~al.} 2016, \mnras, 460,
  1758

\bibitem[{{Hatch} {et~al.}(2007){Hatch}, {Crawford}, \& {Fabian}}]{Hatch2007}
{Hatch}, N.~A., {Crawford}, C.~S., \& {Fabian}, A.~C. 2007, \mnras, 380, 33

\bibitem[{{Hatch} {et~al.}(2006){Hatch}, {Crawford}, {Johnstone}, \&
  {Fabian}}]{Hatch2006}
{Hatch}, N.~A., {Crawford}, C.~S., {Johnstone}, R.~M., \& {Fabian}, A.~C. 2006,
  \mnras, 367, 433

\bibitem[{{Heckman} {et~al.}(1989){Heckman}, {Baum}, {van Breugel}, \&
  {McCarthy}}]{Heckman1989}
{Heckman}, T.~M., {Baum}, S.~A., {van Breugel}, W.~J.~M., \& {McCarthy}, P.
  1989, \apj, 338, 48

\bibitem[{{Hitomi Collaboration}(2016)}]{Hitomi2016}
{Hitomi Collaboration}. 2016, \nat, 535, 117

\bibitem[{{Hogan} {et~al.}(2017){Hogan}, {McNamara}, {Pulido}, {Nulsen},
  {Vantyghem}, {Russell}, {Edge}, {Babyk}, {Main}, \& {McDonald}}]{Hogan2017}
{Hogan}, M.~T., {McNamara}, B.~R., {Pulido}, F.~A., {et~al.} 2017, \apj, 851,
  66

\bibitem[{{Jacob} \& {Pfrommer}(2017{\natexlab{a}})}]{Jacob2017a}
{Jacob}, S. \& {Pfrommer}, C. 2017{\natexlab{a}}, \mnras, 467, 1449

\bibitem[{{Jacob} \& {Pfrommer}(2017{\natexlab{b}})}]{Jacob2017b}
{Jacob}, S. \& {Pfrommer}, C. 2017{\natexlab{b}}, \mnras, 467, 1478

\bibitem[{{Ji} {et~al.}(2018){Ji}, {Oh}, \& {McCourt}}]{Ji2018}
{Ji}, S., {Oh}, S.~P., \& {McCourt}, M. 2018, \mnras, 476, 852

\bibitem[{{Kannan} {et~al.}(2017){Kannan}, {Vogelsberger}, {Pfrommer},
  {Weinberger}, {Springel}, {Hernquist}, {Puchwein}, \&
  {Pakmor}}]{Kannanetal17}
{Kannan}, R., {Vogelsberger}, M., {Pfrommer}, C., {et~al.} 2017, \apj, 837, L18

\bibitem[{{Kimm} {et~al.}(2015){Kimm}, {Cen}, {Devriendt}, {Dubois}, \&
  {Slyz}}]{Kimmetal15}
{Kimm}, T., {Cen}, R., {Devriendt}, J., {Dubois}, Y., \& {Slyz}, A. 2015,
  \mnras, 451, 2900

\bibitem[{{Klein} {et~al.}(1994){Klein}, {McKee}, \& {Colella}}]{Klein1994}
{Klein}, R.~I., {McKee}, C.~F., \& {Colella}, P. 1994, \apj, 420, 213

\bibitem[{{Li} \& {Bryan}(2014{\natexlab{a}})}]{Li&Bryan14a}
{Li}, Y. \& {Bryan}, G.~L. 2014{\natexlab{a}}, \apj, 789, 54

\bibitem[{{Li} \& {Bryan}(2014{\natexlab{b}})}]{Li&Bryan14b}
{Li}, Y. \& {Bryan}, G.~L. 2014{\natexlab{b}}, \apj, 789, 153

\bibitem[{{Li} {et~al.}(2017){Li}, {Ruszkowski}, \& {Bryan}}]{Lietal17}
{Li}, Y., {Ruszkowski}, M., \& {Bryan}, G.~L. 2017, \apj, 847, 106

\bibitem[{{Lim} {et~al.}(2012){Lim}, {Ohyama}, {Chi-Hung}, {Dinh-V-Trung}, \&
  {Shiang-Yu}}]{Lim2012}
{Lim}, J., {Ohyama}, Y., {Chi-Hung}, Y., {Dinh-V-Trung}, \& {Shiang-Yu}, W.
  2012, \apj, 744, 112

\bibitem[{{Lynds}(1970)}]{Lynds1970}
{Lynds}, R. 1970, \apjl, 159

\bibitem[{{Martizzi} {et~al.}(2019){Martizzi}, {Quataert},
  {Faucher-Gigu{\`e}re}, \& {Fielding}}]{Martizzietal19}
{Martizzi}, D., {Quataert}, E., {Faucher-Gigu{\`e}re}, C.-A., \& {Fielding}, D.
  2019, \mnras, 483, 2465

\bibitem[{{McCourt} {et~al.}(2018){McCourt}, {Oh}, {O'Leary}, \&
  {Madigan}}]{McCourtetal18}
{McCourt}, M., {Oh}, S.~P., {O'Leary}, R., \& {Madigan}, A.-M. 2018, \mnras,
  473, 5407

\bibitem[{{McCourt} {et~al.}(2015){McCourt}, {O'Leary}, {Madigan}, \&
  {Quataert}}]{McCourt2015}
{McCourt}, M., {O'Leary}, R.~M., {Madigan}, A.-M., \& {Quataert}, E. 2015,
  \mnras, 449, 2

\bibitem[{{McCourt} {et~al.}(2012){McCourt}, {Sharma}, {Quataert}, \&
  {Parrish}}]{McCourt2011}
{McCourt}, M., {Sharma}, P., {Quataert}, E., \& {Parrish}, I.~J. 2012, \mnras,
  419, 3319

\bibitem[{{McDonald} {et~al.}(2018){McDonald}, {Gaspari}, {McNamara}, \&
  {Tremblay}}]{McDonald2018}
{McDonald}, M., {Gaspari}, M., {McNamara}, B.~R., \& {Tremblay}, G.~R. 2018,
  \apj, 858, 45

\bibitem[{{McDonald} {et~al.}(2012){McDonald}, {Veilleux}, \&
  {Rupke}}]{McDonald2012}
{McDonald}, M., {Veilleux}, S., \& {Rupke}, D.~S.~N. 2012, \apj, 746, 153

\bibitem[{{McDonald} {et~al.}(2010){McDonald}, {Veilleux}, {Rupke}, \&
  {Mushotzky}}]{Mcdonald2010}
{McDonald}, M., {Veilleux}, S., {Rupke}, D.~S.~N., \& {Mushotzky}, R. 2010,
  \apj, 721, 1262

\bibitem[{{McKinley} {et~al.}(2018){McKinley}, {Tingay}, {Carretti}, {Ellis},
  {Bland-Hawthorn}, {Morganti}, {Line}, {McDonald}, {Veilleux}, {Wahl Olsen},
  {Sidonio}, {Ekers}, {Offringa}, {Procopio}, {Pindor}, {Wayth},
  {Hurley-Walker}, {Bernardi}, {Gaensler}, {Haverkorn}, {Kesteven}, {Poppi}, \&
  {Staveley-Smith}}]{McKinley2018}
{McKinley}, B., {Tingay}, S.~J., {Carretti}, E., {et~al.} 2018, \mnras, 474,
  4056

\bibitem[{{McKinney} {et~al.}(2012){McKinney}, {Tchekhovskoy}, \&
  {Blandford}}]{McKinneyetal12}
{McKinney}, J.~C., {Tchekhovskoy}, A., \& {Blandford}, R.~D. 2012, \mnras, 423,
  3083

\bibitem[{McNamara \& Nulsen(2007)}]{McNamara2007}
McNamara, B. \& Nulsen, P. 2007, Annual Review of Astronomy and Astrophysics,
  45, 117

\bibitem[{{Mittal} {et~al.}(2015){Mittal}, {Whelan}, \& {Combes}}]{Mittal2015}
{Mittal}, R., {Whelan}, J.~T., \& {Combes}, F. 2015, \mnras, 450, 2564

\bibitem[{{Nagai} {et~al.}(2019){Nagai}, {Onishi}, {Kawakatu}, {Fujita},
  {Kino}, {Fukazawa}, {Lim}, {Forman}, {Vrtilek}, \& {Nakanishi}}]{Nagai2019}
{Nagai}, H., {Onishi}, K., {Kawakatu}, N., {et~al.} 2019, arXiv e-prints,
  arXiv:1905.06017

\bibitem[{{Narayan} \& {Medvedev}(2001)}]{Narayan01}
{Narayan}, R. \& {Medvedev}, M.~V. 2001, \apj, 562, L129

\bibitem[{{O'Dea} {et~al.}(2008){O'Dea}, {Baum}, {Privon}, {Noel-Storr},
  {Quillen}, {Zufelt}, {Park}, {Edge}, {Russell}, {Fabian}, {Donahue},
  {Sarazin}, {McNamara}, {Bregman}, \& {Egami}}]{Odea2008}
{O'Dea}, C.~P., {Baum}, S.~A., {Privon}, G., {et~al.} 2008, \apj, 681, 1035

\bibitem[{{Ogiya} {et~al.}(2018){Ogiya}, {Biernacki}, {Hahn}, \&
  {Teyssier}}]{Ogiya18}
{Ogiya}, G., {Biernacki}, P., {Hahn}, O., \& {Teyssier}, R. 2018, arXiv
  e-prints, arXiv:1802.02177

\bibitem[{{Olivares} {et~al.}(2019){Olivares}, {Salom{\'e}}, {Combes}, {Hamer},
  {Guillard}, {Lehnert}, {Polles}, {Beckmann}, {Dubois}, {Donahue}, {Edge},
  {Fabian}, {McNamara}, {Rose}, {Russell}, {Tremblay}, {Vantyghem}, {Canning},
  {Ferland }, {Godard}, {Hogan}, {Peirani}, \& {Pineau des
  Forets}}]{Olivares2019}
{Olivares}, V., {Salom{\'e}}, P., {Combes}, F., {et~al.} 2019, arXiv e-prints,
  arXiv:1902.09164

\bibitem[{{Perret} {et~al.}(2014){Perret}, {Renaud}, {Epinat}, {Amram},
  {Bournaud}, {Contini}, {Teyssier}, \& {Lambert}}]{Perretetal14}
{Perret}, V., {Renaud}, F., {Epinat}, B., {et~al.} 2014, \aap, 562, A1

\bibitem[{{Pfister} {et~al.}(2019){Pfister}, {Volonteri}, {Dubois}, {Dotti}, \&
  {Colpi}}]{Pfisteretal19}
{Pfister}, H., {Volonteri}, M., {Dubois}, Y., {Dotti}, M., \& {Colpi}, M. 2019,
  \mnras, 486, 101

\bibitem[{{Pfrommer}(2013)}]{Pfrommer2013}
{Pfrommer}, C. 2013, \apj, 779, 10

\bibitem[{{Prasad} {et~al.}(2015){Prasad}, {Sharma}, \& {Babul}}]{Prasad2015}
{Prasad}, D., {Sharma}, P., \& {Babul}, A. 2015, \apj, 811, 108

\bibitem[{{Prasad} {et~al.}(2018){Prasad}, {Sharma}, \& {Babul}}]{Prasad2018}
{Prasad}, D., {Sharma}, P., \& {Babul}, A. 2018, \apj, 863, 62

\bibitem[{{Pudritz} {et~al.}(2012){Pudritz}, {Hardcastle}, \&
  {Gabuzda}}]{Pudritz2012}
{Pudritz}, R.~E., {Hardcastle}, M.~J., \& {Gabuzda}, D.~C. 2012, \ssr, 169, 27

\bibitem[{{Pulido} {et~al.}(2018){Pulido}, {McNamara}, {Edge}, {Hogan},
  {Vantyghem}, {Russell}, {Nulsen}, {Babyk}, \& {Salom{\'e}}}]{Pulido2018}
{Pulido}, F.~A., {McNamara}, B.~R., {Edge}, A.~C., {et~al.} 2018, \apj, 853,
  177

\bibitem[{{Rafferty} {et~al.}(2006){Rafferty}, {McNamara}, {Nulsen}, \&
  {Wise}}]{Rafferty2006}
{Rafferty}, D.~A., {McNamara}, B.~R., {Nulsen}, P.~E.~J., \& {Wise}, M.~W.
  2006, \apj, 652, 216

\bibitem[{{Reimer} {et~al.}(2004){Reimer}, {Reimer}, {Schlickeiser}, \&
  {Iyudin}}]{Reimer2004}
{Reimer}, A., {Reimer}, O., {Schlickeiser}, R., \& {Iyudin}, A. 2004, \aap,
  424, 773

\bibitem[{{Revaz} {et~al.}(2008){Revaz}, {Combes}, \& {Salom{\'e}}}]{Revaz2008}
{Revaz}, Y., {Combes}, F., \& {Salom{\'e}}, P. 2008, \aap, 477, L33

\bibitem[{{Rosen} \& {Bregman}(1995)}]{Rosen&Bregman95}
{Rosen}, A. \& {Bregman}, J.~N. 1995, \apj, 440, 634

\bibitem[{{Russell} {et~al.}(2017){Russell}, {McDonald}, {McNamara}, {Fabian},
  {Nulsen}, {Bayliss}, {Benson}, {Brodwin}, {Carlstrom}, {Edge},
  {Hlavacek-Larrondo}, {Marrone}, {Reichardt}, \& {Vieira}}]{Russell2016}
{Russell}, H.~R., {McDonald}, M., {McNamara}, B.~R., {et~al.} 2017, \apj, 836,
  130

\bibitem[{{Ruszkowski} \& {Oh}(2010)}]{Ruszkowski11}
{Ruszkowski}, M. \& {Oh}, S.~P. 2010, \apj, 713, 1332

\bibitem[{{Ruszkowski} {et~al.}(2017){Ruszkowski}, {Yang}, \&
  {Reynolds}}]{Ruszkowskietal17}
{Ruszkowski}, M., {Yang}, H.-Y.~K., \& {Reynolds}, C.~S. 2017, \apj, 844, 13

\bibitem[{{Salom{\'e}} {et~al.}(2011){Salom{\'e}}, {Combes}, {Revaz}, {Downes},
  {Edge}, \& {Fabian}}]{Salome2011}
{Salom{\'e}}, P., {Combes}, F., {Revaz}, Y., {et~al.} 2011, \aap, 531, A85

\bibitem[{{Salom{\'e}} {et~al.}(2016){Salom{\'e}}, {Salom{\'e}}, {Combes},
  {Hamer}, \& {Heywood}}]{Salome2016}
{Salom{\'e}}, Q., {Salom{\'e}}, P., {Combes}, F., {Hamer}, S., \& {Heywood}, I.
  2016, \aap, 586, A45

\bibitem[{{Salom\'e, P.} {et~al.}(2006){Salom\'e, P.}, {Combes, F.}, {Edge, A.
  C.}, {Crawford, C.}, {Erlund, M.}, {Fabian, A. C.}, {Hatch, N. A.},
  {Johnstone, R. M.}, {Sanders, J. S.}, \& {Wilman, R. J.}}]{Salome2006}
{Salom\'e, P.}, {Combes, F.}, {Edge, A. C.}, {et~al.} 2006, A\&A, 454, 437

\bibitem[{{Salom\'e, P.} {et~al.}(2008){Salom\'e, P.}, {Combes, F.}, {Revaz,
  Y.}, {Edge, A. C.}, {Hatch, N. A.}, {Fabian, A. C.}, \& {Johnstone, R.
  M.}}]{Salome2008}
{Salom\'e, P.}, {Combes, F.}, {Revaz, Y.}, {et~al.} 2008, A\&A, 484, 317

\bibitem[{{Schmidt} {et~al.}(2002){Schmidt}, {Fabian}, \&
  {Sanders}}]{Schmidt2002}
{Schmidt}, R.~W., {Fabian}, A.~C., \& {Sanders}, J.~S. 2002, \mnras, 337, 71

\bibitem[{{Segers} {et~al.}(2017){Segers}, {Oppenheimer}, {Schaye}, \&
  {Richings}}]{Segersetal17}
{Segers}, M.~C., {Oppenheimer}, B.~D., {Schaye}, J., \& {Richings}, A.~J. 2017,
  \mnras, 471, 1026

\bibitem[{{Sharma} {et~al.}(2012){Sharma}, {McCourt}, {Quataert}, \&
  {Parrish}}]{Sharma2012}
{Sharma}, P., {McCourt}, M., {Quataert}, E., \& {Parrish}, I.~J. 2012, \mnras,
  420, 3174

\bibitem[{{Shi} {et~al.}(2011){Shi}, {Helou}, {Yan}, {Armus}, {Wu}, {Papovich},
  \& {Stierwalt}}]{Shi2011}
{Shi}, Y., {Helou}, G., {Yan}, L., {et~al.} 2011, \apj, 733, 87

\bibitem[{{Shin} {et~al.}(2008){Shin}, {Stone}, \& {Snyder}}]{Shin08}
{Shin}, M.-S., {Stone}, J.~M., \& {Snyder}, G.~F. 2008, \apj, 680, 336

\bibitem[{{Sparre} {et~al.}(2019){Sparre}, {Pfrommer}, \&
  {Vogelsberger}}]{Sparre2019}
{Sparre}, M., {Pfrommer}, C., \& {Vogelsberger}, M. 2019, \mnras, 482, 5401

\bibitem[{{Sutherland} \& {Dopita}(1993)}]{Sutherland&Dopita93}
{Sutherland}, R.~S. \& {Dopita}, M.~A. 1993, \apjs, 88, 253

\bibitem[{{Teyssier}(2002)}]{Teyssier02}
{Teyssier}, R. 2002, \aap, 385, 337

\bibitem[{{Tremblay} {et~al.}(2018){Tremblay}, {Combes}, {Oonk}, {Russell},
  {McDonald}, {Gaspari}, {Husemann}, {Nulsen}, {McNamara}, \&
  {Hamer}}]{Tremblay2018}
{Tremblay}, G.~R., {Combes}, F., {Oonk}, J.~B.~R., {et~al.} 2018, \apj, 865, 13

\bibitem[{{Turk} {et~al.}(2011){Turk}, {Smith}, {Oishi}, {Skory}, {Skillman},
  {Abel}, \& {Norman}}]{Turk2011}
{Turk}, M.~J., {Smith}, B.~D., {Oishi}, J.~S., {et~al.} 2011, The Astrophysical
  Journal Supplement Series, 192, 9

\bibitem[{{Vantyghem} {et~al.}(2017){Vantyghem}, {McNamara}, {Edge}, {Combes},
  {Russell}, {Fabian}, {Hogan}, {McDonald}, {Nulsen}, \&
  {Salom{\'e}}}]{Vantyghem2017}
{Vantyghem}, A.~N., {McNamara}, B.~R., {Edge}, A.~C., {et~al.} 2017, \apj, 848,
  101

\bibitem[{{Vantyghem} {et~al.}(2018){Vantyghem}, {McNamara}, {Russell}, {Edge},
  {Nulsen}, {Combes}, {Fabian}, {McDonald}, \& {Salom{\'e}}}]{Vantyghem2018}
{Vantyghem}, A.~N., {McNamara}, B.~R., {Russell}, H.~R., {et~al.} 2018, \apj,
  863, 193

\bibitem[{{Voit}(2018)}]{Voit2018}
{Voit}, G.~M. 2018, \apj, 868, 102

\bibitem[{{Voit} \& {Donahue}(2015)}]{Voit2015}
{Voit}, G.~M. \& {Donahue}, M. 2015, \apj, 799, L1

\bibitem[{{Voit} {et~al.}(2017){Voit}, {Meece}, {Li}, {O'Shea}, {Bryan}, \&
  {Donahue}}]{Voit2017}
{Voit}, G.~M., {Meece}, G., {Li}, Y., {et~al.} 2017, \apj, 845, 80

\bibitem[{{Wang} {et~al.}(2019){Wang}, {Li}, \& {Ruszkowski}}]{Wangetal19}
{Wang}, C., {Li}, Y., \& {Ruszkowski}, M. 2019, \mnras, 482, 3576

\bibitem[{{Werner} {et~al.}(2013){Werner}, {Urban}, {Simionescu}, \&
  {Allen}}]{Werner2013}
{Werner}, N., {Urban}, O., {Simionescu}, A., \& {Allen}, S.~W. 2013, \nat, 502,
  656

\bibitem[{{Xu} \& {Lazarian}(2018)}]{Xu2019}
{Xu}, S. \& {Lazarian}, A. 2018, arXiv e-prints, arXiv:1802.00987

\bibitem[{{Yang} \& {Reynolds}(2016{\natexlab{a}})}]{Yang&Reynolds16}
{Yang}, H.-Y.~K. \& {Reynolds}, C.~S. 2016{\natexlab{a}}, \apj, 829, 90

\bibitem[{{Yang} \& {Reynolds}(2016{\natexlab{b}})}]{Yang&Reynolds16Cond}
{Yang}, H. Y.~K. \& {Reynolds}, C.~S. 2016{\natexlab{b}}, \apj, 818, 181

\end{thebibliography}

%%%%%%%%%%%%%%%%%%%%%%%%%%%%%%%%%%%%%%%%%%%%%%%%%%

%%%%%%%%%%%%%%%%% APPENDICES %%%%%%%%%%%%%%%%%%%%%

%\appendix



%%%%%%%%%%%%%%%%%%%%%%%%%%%%%%%%%%%%%%%%%%%%%%%%%%

% Don't change these lines

\end{document}